\documentclass[iop,apj]{emulateapj}
\usepackage{apjfonts}
\usepackage{natbib}
\usepackage{amsfonts}
\usepackage{amsmath}
\usepackage{multirow}
\usepackage{enumerate}
\usepackage[normalem]{ulem}
\usepackage[usenames,dvipsnames]{xcolor}
\def\sun{_\odot}

\def\Dwa{$\,$\uppercase\expandafter{\romannumeral5}$\,$}

\def\sless{\lower2pt\hbox{$\buildrel {\scriptstyle <}
   \over {\scriptstyle\sim}$}}

\def\sgreat{\lower2pt\hbox{$\buildrel {\scriptstyle >}
   \over {\scriptstyle\sim}$}}
\def\sharpnull#1{}

\def\cqg{Class.\ Quantum Grav.}

\newcommand{\numberofeos}{8}

\begin{document}
\slugcomment{Submitted to ApJ. June 17, 2013. Accepted for publication on June 1, 2014.}

\title{The Influence of Thermal Pressure on Equilibrium Models \\ of
  Hypermassive Neutron Star Merger Remnants}

\author{J. D. Kaplan\altaffilmark{1}, C. D. Ott\altaffilmark{1,2,$\star$}, E. P. O'Connor\altaffilmark{3}, K. Kiuchi\altaffilmark{4}, L. Roberts\altaffilmark{1}, and M. Duez\altaffilmark{5} }

\altaffiltext{1}{TAPIR, Mailcode 350-17,
California Institute of Technology, Pasadena, CA 91125, USA;
cott@tapir.caltech.edu} 
\altaffiltext{2}{Kavli IPMU (WPI), University of Tokyo, Kashiwa, Japan}
\altaffiltext{3}{CITA, 60 St. George Street, University of Toronto, Toronto, Canada M5S 3H8}
\altaffiltext{4}{Yukawa Institute for Theoretical Physics, University of Kyoto, Kyoto, Japan}
\altaffiltext{5}{Department of Physics and Astronomy, Washington State University, Pullman, WA, USA}
\altaffiltext{$\star$}{
Alfred P. Sloan Research Fellow}

\begin{abstract}
The merger of two neutron stars leaves behind a rapidly spinning
hypermassive object whose survival is believed to depend on the
maximum mass supported by the nuclear equation of state, angular
momentum redistribution by (magneto-)rotational instabilities, and
spindown by gravitational waves. The high temperatures
($\sim$$5-40\,\mathrm{MeV}$) prevailing in the merger remnant may
provide thermal pressure support that could increase its maximum mass
and, thus, its life on a neutrino-cooling timescale.  We investigate
the role of thermal pressure support in hypermassive merger remnants
by computing sequences of spherically-symmetric and axisymmetric
uniformly and differentially rotating equilibrium solutions to the
general-relativistic stellar structure equations.  Using a set of
finite-temperature nuclear equations of state, we find that hot
maximum-mass critically spinning configurations generally do not
support larger baryonic masses than their cold counterparts. However,
subcritically spinning configurations with mean density of less than a
few times nuclear saturation density yield a significantly thermally
enhanced mass.  Even without decreasing the maximum mass, cooling and
other forms of energy loss can drive the remnant to an unstable state.
We infer secular instability by identifying approximate energy turning
points in equilibrium sequences of constant baryonic mass parametrized
by maximum density.  Energy loss carries the remnant along the
direction of decreasing gravitational mass and higher density until
instability triggers collapse.  Since configurations with more thermal
pressure support are less compact and thus begin their evolution at a
lower maximum density, they remain stable for longer periods after
merger.
\end{abstract}


\section{Introduction}
\label{sec:Introduction}

Coalescing double neutron stars (NSs) are prime candidate progenitors
of short-hard gamma-ray bursts (GRBs, e.g., \citealt{nakar:07a} and
references therein). The strong gravitational wave emission driving
the coalescence makes NSNS systems the primary targets of the network
of second-generation gravitational-wave interferometers currently
under construction (Advanced LIGO [\citealt{harry:10}], Advanced Virgo
[\citealt{virgostatus:11}], and KAGRA [\citealt{kagra:12}]).

Until the last moments of inspiral, the constituent NSs may
essentially be treated as cold neutron stars. Tidal heating is mild
and the NS crust may not fail until the NSs touch
(\citealt{penner:12}, but see \citealt{tsang:12} and
\citealt{weinberg:13}). Merger results in the formation of a shocked,
extremely rapidly differentially spinning central object, commonly
referred to as a hypermassive NS (HMNS), since it comprises the vast
majority of the baryonic mass of the two premerger NSs and is thus
expected to be more massive than the maximum mass supported by the
nuclear equation of state (EOS) in the spherical and uniformly
rotating limits limit (see, e.g., \citealt{faber:12} for a review of
NSNS mergers). The subsequent evolution of the HMNS has important
ramifications for gravitational wave emission and the possible
transition to a short-hard GRB. If the HMNS survives for an extended
period, nonaxisymmetric rotational instability may enhance the
high-frequency gravitational-wave emission, possibly allowing
gravitational-wave observers to constrain the nuclear EOS (e.g.,
\citealt{bauswein:12}). On the other hand, the neutrino-driven wind
blown off a surviving HMNS, producing mass loss at a rate of order
$10^{-4}\,M_\odot\,\mathrm{s}^{-1}$, will lead to strong baryon
loading in polar regions \citep{dessart:09}, making the formation of
the relativistic outflows needed for a GRB more difficult, even if a
black hole with an accretion disk forms eventually. If the HMNS
collapses to a black hole within milliseconds of merger, baryon
loading will not hamper a GRB, but strong gravitational-wave and
neutrino emission would be shut off rapidly.

The long-term survival of the HMNS depends sensitively on the maximum
mass of a nonrotating cold neutron star supported by the nuclear EOS,
which most certainly is above $\sim 2\,M_\odot$
\citep{demorest:10,antoniadis:13} and very likely below $\sim
3.2\,M_\odot$ \citep{lattimer:07}. At its formation, the HMNS is
rapidly and strongly differentially rotating.  Extreme differential
rotation alone may increase the maximum HMNS mass by more than $100\%$
\citep[e.g.,][]{baumgarte:00}. Angular momentum redistribution by
(magneto-)rotational instabilities and spindown by gravitational wave
emission are expected to remove this additional support. This will
ultimately lead to black hole formation if the HMNS mass is above the
maximum mass that can be supported by the nuclear EOS and uniform
rotation ($\lesssim$$20\%$ greater than the maximum in the nonrotating
limit; \citealt{baumgarte:00}).

Recently, \cite{sekiguchi:11nsns}, \cite{paschalidis:12},
\cite{bauswein:10}, and, in earlier work, \cite{baiotti:08a}, have
argued that thermal pressure support at moderately high temperatures
of $\sim 5 - 40\,\mathrm{MeV}$ \citep{oechslin:07,sekiguchi:11nsns}
may significantly influence the structure and evolution of the
postmerger HMNS and prolong its lifetime until collapse to a black
hole. If true, the HMNS may survive on the neutrino cooling timescale
provided that the combined premerger mass of the NSs is sufficiently
close to the thermally-enhanced maximum HMNS mass.  These authors
estimate the neutrino cooling timescale to be comparable to or longer
than the timescale for angular momentum redistribution and spindown by
gravitational waves.

The focus of this paper is on the role of thermal pressure support in
hypermassive NS merger remnants.  Postmerger HMNS configurations that
survive for multiple dynamical times quickly assume dynamical
equilibrium and, after the extremely dynamic merger phase, show only
mild deviation from axisymmetry (e.g.,
\citealt{sekiguchi:11nsns,shibata:2005ss}). Hence, instead of
performing computationally expensive full merger simulations, we
investigate the role of thermal effects by approximating HMNS
configurations as sequences of rotational equilibrium solutions, which
we compute with the relativistic self-consistent field method
\citep{komatsu:89a,komatsu:89b, cook:92}. We consider the spherical
limit (Tolman-Oppenheimer-Volkoff [TOV] solutions), uniform, and
differential rotation. We employ multiple finite-temperature
microphysical nuclear EOS and, since the equilibrium solver requires a
barotropic equation of state, a range of temperature and composition
parametrizations that are motivated by the merger simulations of
\cite{sekiguchi:11nsns}. An overall similar approach, though only
considering isothermal and isentropic configurations, has been used in
the past to study thermal effects on uniformly and differentially
rotating proto-neutron stars \citep{goussard:97,goussard:98}.

The key quantity relevant in the secular evolution of HMNSs is the
baryonic mass ($M_\mathrm{b}$; also called ``rest mass'') that can be
supported by a given combination of EOS, thermal/compositional
structure, and rotational setup. The gravitational mass
($M_\mathrm{g}$) is not conserved and is reduced by cooling and
angular momentum loss.  \emph{Our results show that the maximum
  baryonic mass of TOV, uniformly rotating, and differentially
  rotating configurations is essentially unaffected by thermal
  pressure support}.  Thermal pressure support is negligible at
supranuclear densities and becomes significant only at densities below
nuclear saturation density. Since maximum-mass configurations always
have maximum and mean densities above nuclear, thermal pressure
support is minimal.  The thermal contribution to the stress-energy
tensor (which sources curvature) may, depending on the EOS, even lead
to a net decrease of the $M_\mathrm{b}^\mathrm{max}$ with increasing
temperature.

We find thermal enhancement of $M_\mathrm{b}$ for configurations with
mean densities less than a few times nuclear saturation density that
are nonrotating or rotating subcritically (i.e., below the
mass-shedding limit). A hot configuration in this regime will support
the same baryonic mass at a lower mean (and maximum) density.
However, hot rotating configurations are spatially more extended than
their cold counterparts, and thus reach mass shedding at lower angular
velocities.  This counteracts the thermal enhancement and results in
$M_\mathrm{b}^\mathrm{max}$ that are within a few percent of cold
configurations.

The secular evolution of a HMNS towards collapse is driven by energy
losses to gravitational waves and neutrinos, and, potentially, by loss
of angular momentum transported to the surface by processes such as
the MRI.  It proceeds along trajectories of constant (or nearly
constant) baryonic mass and in the direction of decreasing total
energy (i.e., gravitational mass $M_\mathrm{g}$) and increasing
maximum baryon density $\rho_\mathrm{b,max}$ (i.e., more compact
configurations). We conjecture, based on established results of the
theory of rotating relativistic stars (\citealt{friedman:13}), that
instability to collapse occurs when the configuration reaches an
unstable part of the parameter space and not necessarily because the
maximum supportable baryonic mass $M_\mathrm{b}^\mathrm{max}$ drops
below $M_\mathrm{b}$. We formalize this via an approximate variant of
the turning-point theorem (e.g., \citealt{sorkin:82,friedman:13}): The
turning-point theorem states that for uniformly rotating neutron
stars, a local extremum in $M_\mathrm{g}$
at fixed angular momentum, entropy, and baryonic mass constitutes a
point at which secular instability to collapse must set in. We argue
that the turning point theorem carries over to differentially rotating
hot HMNSs. The precise turning points become approximate and are
distributed over a narrow range of $\rho_\mathrm{b,max}$ and
$M_\mathrm{g}$ for all degrees of differential rotation and
temperature prescriptions that we consider here.  The regime of
instability is thus largely independent of HMNS temperature.  However,
a hotter configuration will be less compact initially and, hence, will
begin its secular evolution to its turning point at a lower
$\rho_\mathrm{b,max}$ than a colder one. It will thus have to evolve
further until it reaches its turning point and, at a fixed rate of
energy loss, will survive for longer.

This paper is structured as follows.  In \S\ref{sec:methods}, we
introduce the set of EOS we employ and discuss the relative importance
of thermal pressure as a function of density. We also introduce the
temperature and composition parametrizations and the methods used for
constructing equilibrium models without and with rotation. In
\S\ref{sec:TOVresults}, we lay out our results for nonrotating NSs and
then discuss uniformly and differentially rotating configurations in
\S\ref{sec:results_2Duni} and \S\ref{sec:results_2Ddiff},
respectively. We consider evolutionary sequences of HMNSs at constant
baryonic mass in the context of an approximate turning point theorem
and compare with results from recent merger simulations in
\S\ref{sec:compare}.  Finally, in \S\ref{sec:conclusions}, we
summarize our results and conclude.

\section{Methods and Equations of State}
\label{sec:methods}
\subsection{Equations of State}
\label{sec:methods:eos}

\begin{figure}[t]
\centering
\includegraphics[width=\columnwidth]{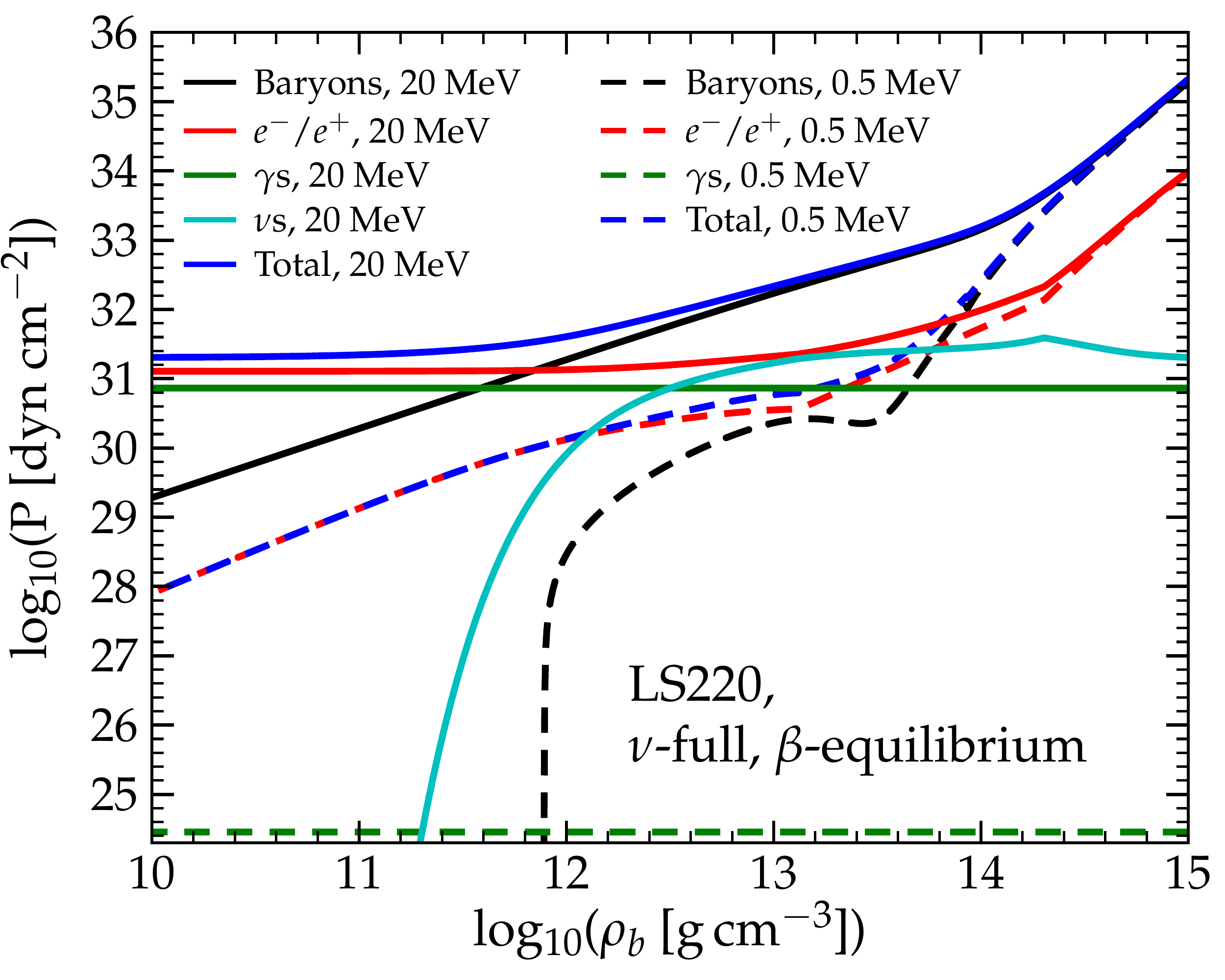}
\caption{Individual pressure contributions of baryons,
  electrons/positrons, photons, and trapped neutrinos and the total
  pressure as a function of baryon density in the LS220 EOS for
  $\nu$-full $\beta$-equilibrium as described in the text and $T =
  0.5\,\mathrm{MeV}$ (dashed lines) and $T=20\,\mathrm{MeV}$ (solid
  lines). The qualitative and quantitative behavior of the LS220 EOS
  with increasing temperature is representative for all EOS considered
  in this study. Note that the baryon pressure becomes negative at
  $\rho_\mathrm{b} \lesssim 10^{12}\,\mathrm{g\,cm}^{-3}$, and dips around
  $10^{13.5}\,\mathrm{g\,cm}^{-3}$ due to Coulomb effects at low
  temperatures \citep{lseos:91}.}
\label{fig:pressure_contributions}
\end{figure}

We use a set of \numberofeos\ EOS in this study. All EOS produce cold
neutron stars in $\beta$-equilibrium that can have gravitational
masses $M_\mathrm{g}$ above $2\,M_\odot$. These include two EOS from
\cite{lseos:91}, the $K_0=220\,$MeV and $K_0=375\,$MeV variants (where
$K_0$ is the nuclear compressibility modulus), denoted LS220 and
LS375; the relativistic mean field (RMF) model EOS from
\cite{hshen:11}, denoted HShen; two RMF models based on the NL3 and
the FSUGold parameter set \citep{gshen:11a,gshen:11b} denoted
GShen-NL3 and GShen-FSU2.1; an unpublished\footnote{Available from
  \url{http://phys-merger.physik.unibas.ch/\textasciitilde
    hempel/eos.html}, based on \cite{hempel:12,hempel:10}.} RMF model
based on the DD2 interaction denoted HSDD2; and two recent RMF model
EOS fit to astrophysical measurements of neutron star masses and radii
\citep{steiner:13b}, denoted SFHo and SFHx. All of these EOS are
available in a common format for download from {\tt
  \url{http://www.stellarcollapse.org}}.

The EOS of finite-temperature nuclear matter in nuclear statistical
equilibrium (NSE) has contributions from a baryonic component
(nucleons and nuclei), a relativistic electron/positron Fermi gas, a
photon gas, and, if neutrinos are trapped, a neutrino gas.  The
Helmholtz free energies of these components add linearly, and the
pressure is then the sum of the partial pressures and a function of
baryon density $\rho$, temperature $T$ and electron fraction
$Y_e$,
\begin{equation}
P = P_\mathrm{baryon} + P_e + P_\gamma + P_\nu\,\,.
\end{equation}
While $P_\mathrm{baryon}$ varies between the employed EOS, we add
$P_e$ and $P_\gamma$ using the Timmes EOS \citep{timmes:99} available
from {\tt \url{http://cococubed.asu.edu}}.  In hot HMNSs, like in
protoneutron stars, neutrinos are trapped and in equilibrium with
matter. We include their pressure contribution to the EOS by treating
them as a non-interacting relativistic Fermi gas with chemical
potential $\mu_{\nu_i}$.  For a single species of neutrinos and
antineutrinos, the neutrino pressure in equilibrium is
\begin{equation}
P_{\nu_i} = \frac{4\pi (k_B T)^4}{3(hc)^3} \left[F_3\left(\eta_{\nu_i}\right) + F_3\left(-\eta_{\nu_i}\right)\right] \times \exp{\left(-\frac{\rho_\mathrm{trap}}{\rho}\right)}\,\,,
\end{equation}
where $\eta_{\nu_i} = \mu_{\nu_i} / (k_B T)$ is the neutrino
degeneracy parameter. For HMNS conditions, all neutrino species are
present, but $\nu_\mu$ and $\nu_\tau$ have $\mu_{\nu_i}=0$, since they
appear only in particle--anti-particle pairs that have equal and
opposite chemical potentials. For electron neutrinos we use
$\mu_{\nu_e} = \mu_e + \mu_p - \mu_n$, for electron antineutrinos we
use $\mu_{\bar{\nu}_e} = -\mu_{\nu_e}$.  We include an attenuation
factor $\exp(-\rho_\mathrm{trap}/\rho)$ to account for the fact that
neutrinos decouple from matter at low densities. We set
$\rho_\mathrm{trap} = 10^{12.5}\,\mathrm{g}\,\mathrm{cm^{-3}}$, which
is a fiducial trapping density for protoneutron stars (e.g.,
\citealt{liebendoerfer:05fakenu}). Taking the exact expression for the
difference of the Fermi integrals from \cite{bludman:78}, we have the
total neutrino pressure summed over all three species,
\begin{equation}
P_\nu = \frac{4\pi (k_B T)^4}{3(hc)^3} \left[\frac{21\pi^4}{60} + \frac{1}{2}\eta_{\nu_e}^2\left(\pi^2 + \frac{1}{2}\eta_{\nu_e}^2\right)\right]  \times \exp{\left(-\frac{\rho_\mathrm{trap}}{\rho}\right)}\,\,.\label{eq:pressure_nu}
\end{equation}

We note that due to the neutrino statistical weight $g = 1$, for a
single species of relativistic non-degenerate $\nu-\bar{\nu}$ pairs,
the pressure is a factor of two lower than for $e^- - e^+$ pairs,
since $e^-$ and $e^+$ have statistical weight (spin degeneracy) $2$.

Figure~\ref{fig:pressure_contributions} illustrates the contributions
of the partial pressures to the total pressure as a function of
baryon density $\rho_\mathrm{b}$ for neutron-rich HMNS matter at two
temperatures, $0.5\,\mathrm{MeV}$ (a representative ``cold''
temperature) and $20\,\mathrm{MeV}$ (a representative ``hot''
temperature for HMNSs). For the 0.5\,MeV EOS, we set the 
electron fraction $Y_e$ by solving for $\nu$-less $\beta$-equilibrium
($\mu_{\nu_e} = 0$).  The resulting EOS describes ordinary cold neutron
stars (at $0.5\,\mathrm{MeV}$ any thermal effects are negligible).
For the $20\,\mathrm{MeV}$ case, we solve for $Y_e$ by assuming
$\nu$-full $\beta$-equilibrium. We do so by making the assumption that
any neutrinos produced during the merger are immediately trapped in
the HMNS core, but stream away from regions below trapping density.
The procedure is discussed in the next section \ref{sec:tempcomp}
and detailed in Appendix~\ref{sec:ye}.

Near and above nuclear saturation density, $\rho_\mathrm{nuc} \simeq
2.6\times10^{14}\,\mathrm{g\,cm}^{-3}$ for the LS220 EOS, the baryon
pressure is due to the repulsive core of the nuclear force and
dominates in both cold and hot regimes. The thermal enhancement above
$\rho_\mathrm{nuc}$ remains small even at $20\,\mathrm{MeV}$. In the
cold case, relativistically degenerate electrons ($\Gamma = (d\ln P)(d
\ln \rho)^{-1} = 4/3$) dominate below $\rho_\mathrm{nuc}$. At 20\,MeV,
relativistic non-degenerate electron/positron pairs and photons (for
both, $P \propto T^4$, independent of $\rho_\mathrm{b}$; see, e.g.,
\citealt{vanriper:77}) are the primary contributors at low densities,
while the baryon pressure is significantly thermally enhanced below
nuclear saturation density and dominates above
$\sim$$10^{12}\,\mathrm{g\,cm}^{-3}$. The neutrino pressure is
comparable to the degenerate electron pressure between
$\sim$$10^{12.5}-10^{14}\,\mathrm{g\,cm}^{-3}$, but still subdominant
to the nuclear component.  The contribution of pairs and photons
gradually becomes more important at all densities as the temperature
increases. We note that for $T=0.5$\,MeV, the neutrino chemical
potentials are all zero and the pressure of trapped neutrinos is
$3\times(7/8)\times P_\gamma$, thermodynamically insignificant at
$T=0.5$\,MeV.

\begin{figure*}[t]
\centering
\includegraphics[width=\columnwidth]{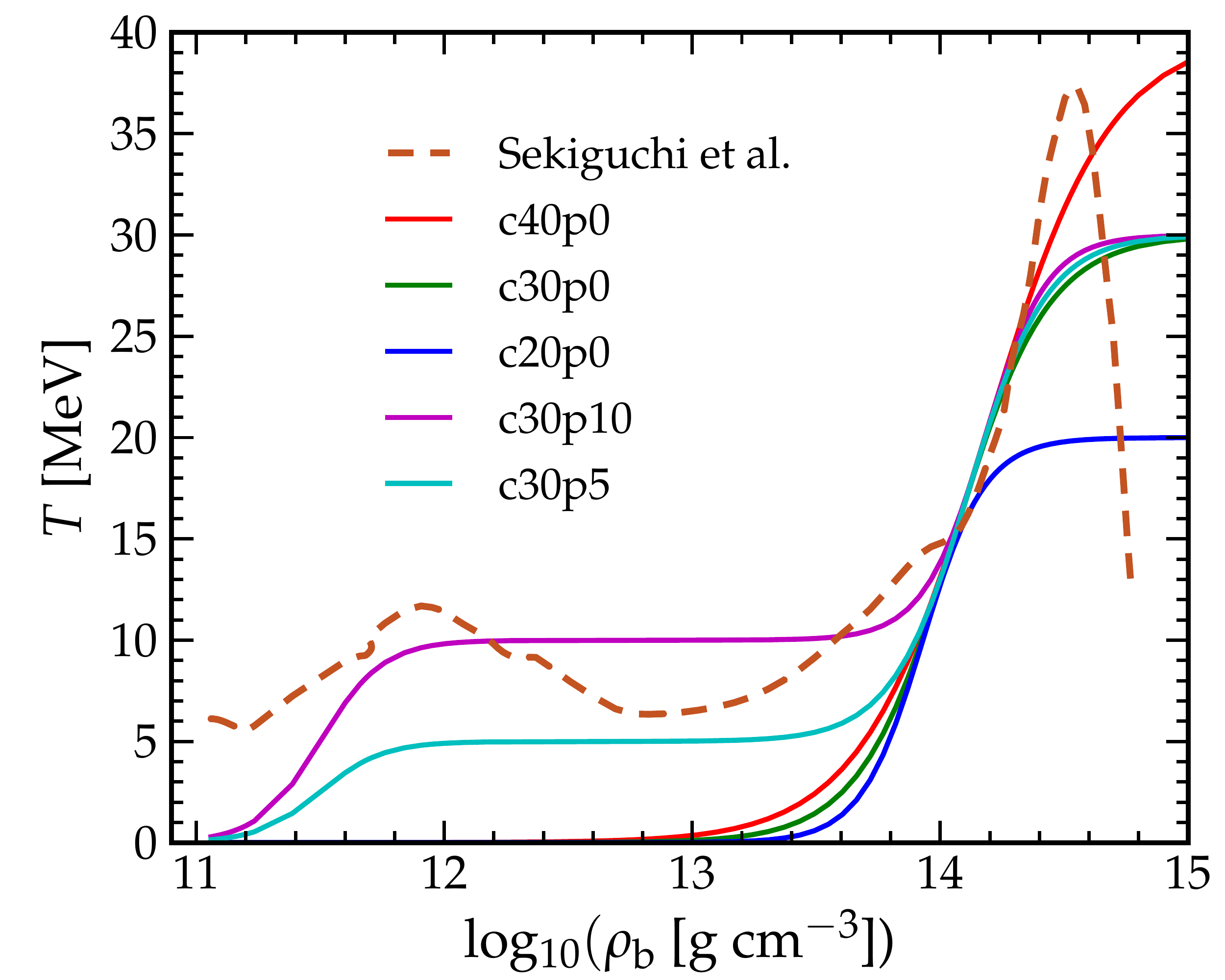}
\includegraphics[width=\columnwidth]{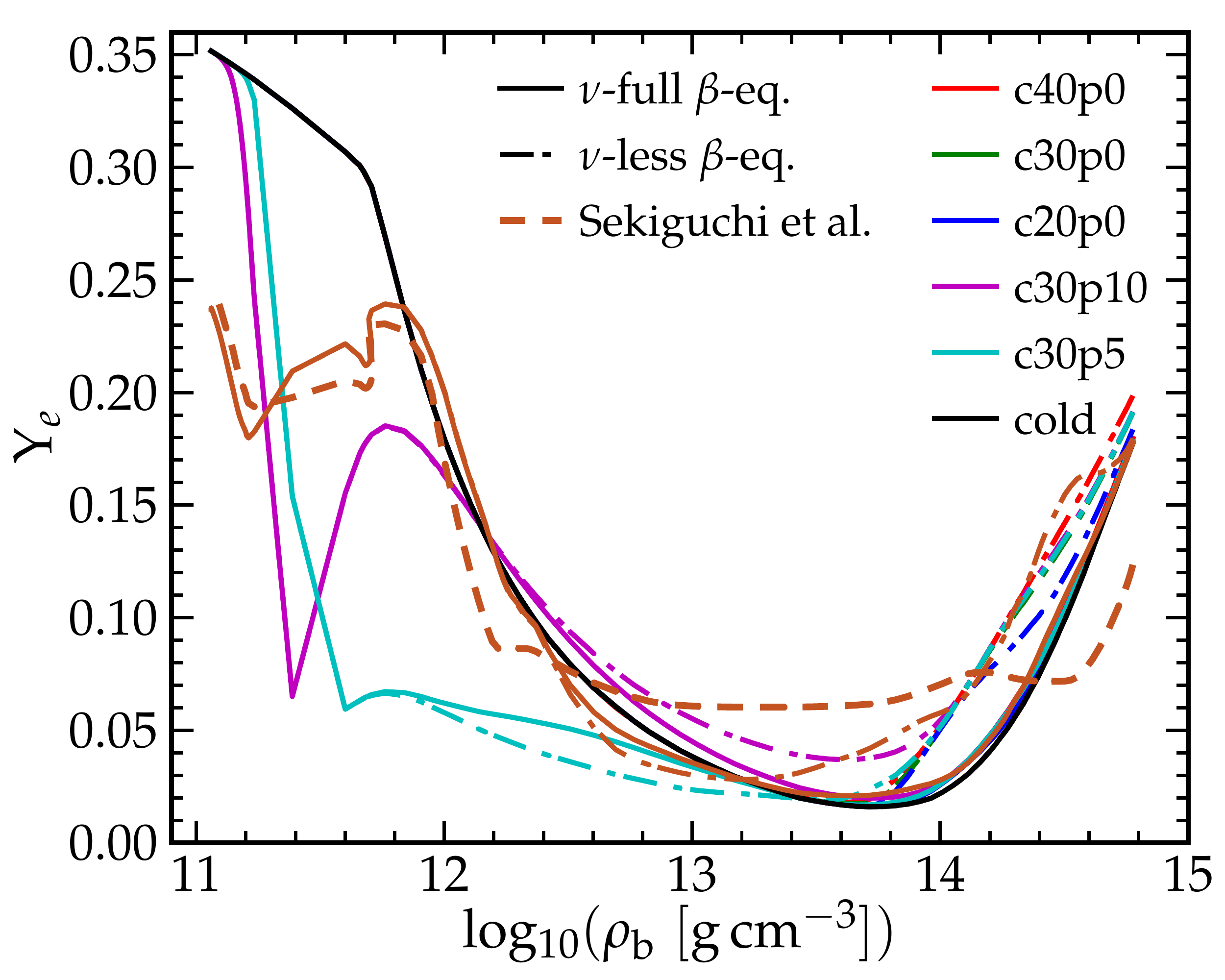}
\caption{Temperature ($T$, left panel) and electron fraction ($Y_e$,
  right panel) as a function of baryon density for the $T$ and $Y_e$
  prescriptions we explore in this work compared to 3D NSNS simulation
  data of \cite{sekiguchi:11nsns} (dashed brown graphs). The profiles
  are created by taking $T$, $Y_e$, and $\rho_b$ data along the
  $+x$-axis from their low-mass (two $1.35$-$M_\odot$ progenitor NSs)
  simulation at $12.1\,\mathrm{ms}$ after merger. In the right panel,
  the dashed brown graph denotes the $Y_e$ obtained from the
  simulation, while the solid and the dash-dotted graphs are $Y_e$
  obtained from the simulation temperature profile for $\nu$-full and
  $\nu$-less $\beta$-equilibrium, respectively.}
\label{fig:tempScripts}
\end{figure*}

\begin{figure*}[t]
\centering
\includegraphics[width=8.5cm]{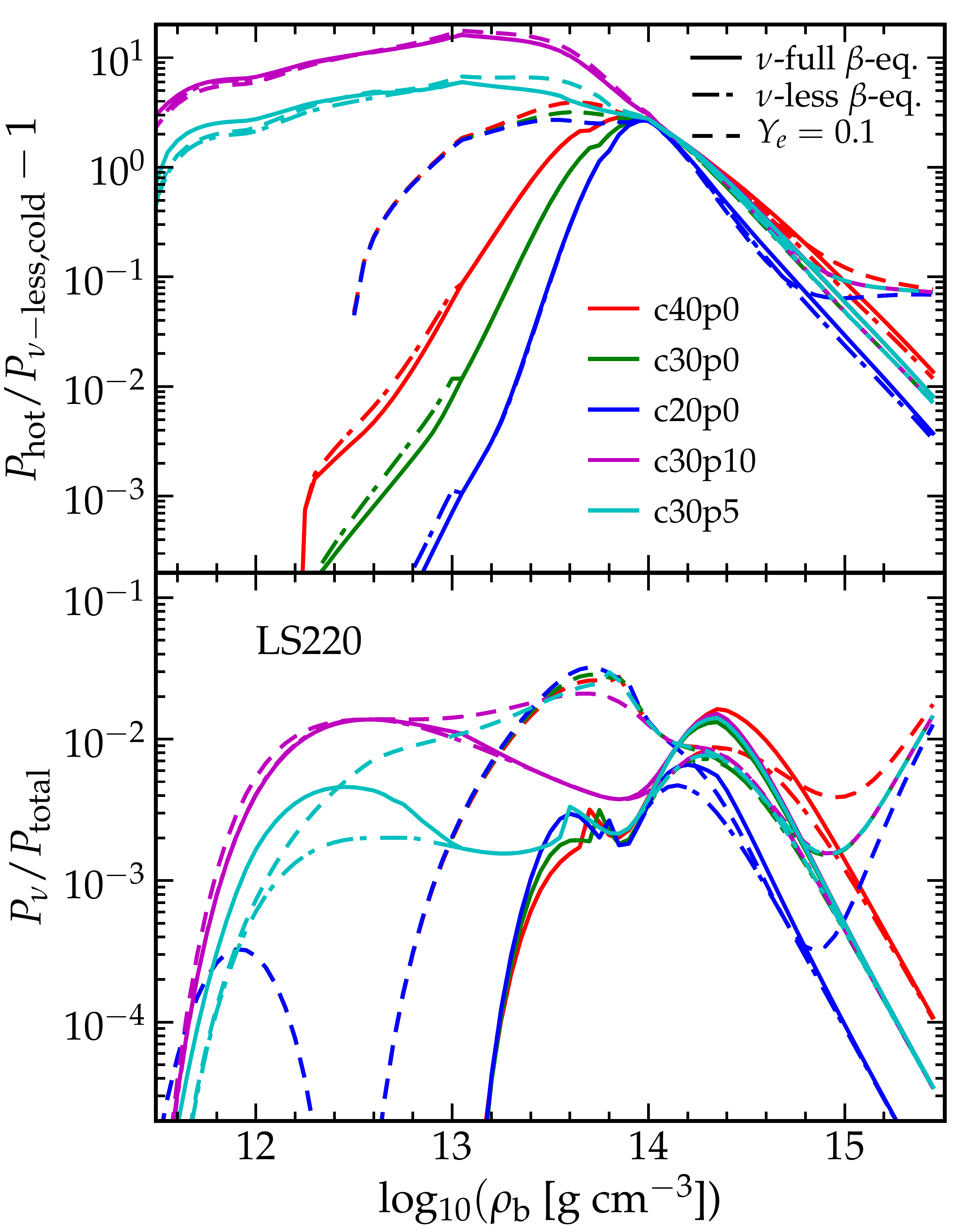}
\hspace*{0.3cm}
\includegraphics[width=8.5cm]{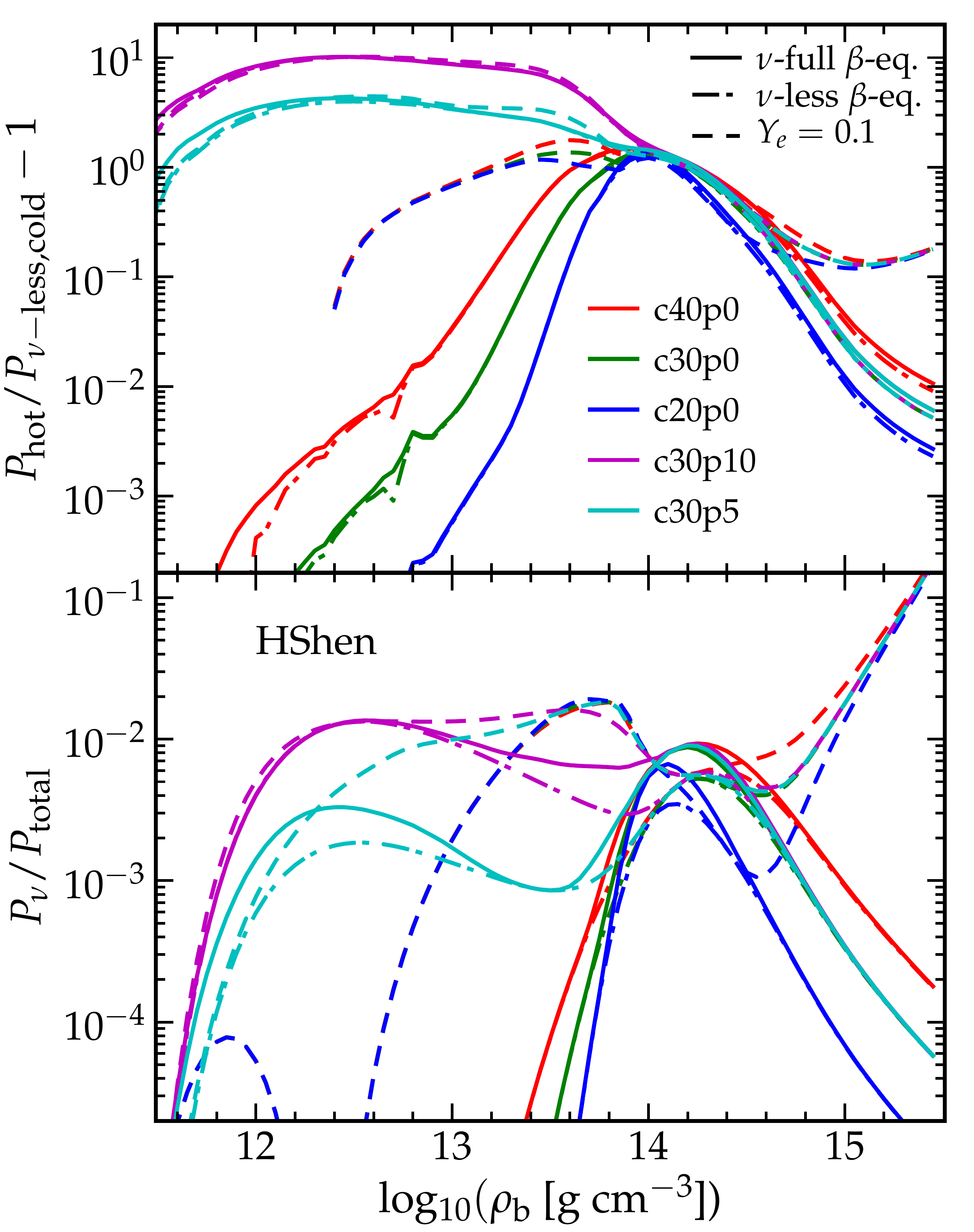}
\caption{Effects of temperature and $Y_e$ parametrizations on the
  pressure and relevance of the neutrino pressure component.  {\bf Top
    panels:} Fractional increase of the pressure over the cold
  $\nu$-less $\beta$-equilibrium pressure for the LS220 EOS (left
  panel) and the HShen EOS (right panel). The different line styles
  correspond to $Y_e(\rho)$ obtained in $\nu$-full $\beta$-equilibrium
  (solid), $\nu$-less $\beta$-equilibrium (dash-dotted), and constant
  $Y_e = 0.1$ (dashed).  {\bf Bottom panels: } Relative contribution of
  the neutrinos to the total pressure (cf. Eq.~\ref{eq:pressure_nu})
  in the five temperature and three $Y_e$ parametrizations and the
  LS220 EOS (left panel) and the HShen EOS (right panel).}
\label{fig:pRatio}
\end{figure*}

\subsection{Temperature and Composition Parametrizations}
\label{sec:tempcomp}

The hydrostatic and rotational equilibrium equations that we solve in
this study assume a barotropic EOS ($P=P(\rho)$) and do not provide
constraints on thermal structure and composition ($Y_e$ is the only
relevant compositional variable in NSE). We must make some assumptions
to be able to proceed and obtain $P=P(\rho,T(\rho),Y_e(\rho,T(\rho)))$ for our
general finite-temperature microphysical EOS. Old NSs in isolation are
nearly isothermal (e.g., \citealt{prakash:01}) and so are coalescing
neutron stars until tidal heating becomes significant (e.g.,
\citealt{kochanek:92,lai:94}). During merger, the NS matter is
shock-heated to tens of MeV and results of the few merger simulations
that have been carried out with temperature-dependent EOS (e.g.,
\citealt{sekiguchi:11nsns,bauswein:10,oechslin:07,rosswog:03b,ruffert:01})
indicate that the HMNS is far from being isothermal or isentropic. It
has a very hot dense core with $T \sim 20 - 40\,\mathrm{MeV}$
surrounded by a lower-density cooler envelope/torus of
$5-20\,\mathrm{MeV}$, which may also be almost Keplerian and, hence,
centrifugally supported. This result appears to be robust for
equal-mass or near equal-mass NSNS systems (which may dominate the
population; e.g., \citealt{lattimer:12} and references therein).
Mergers of non-equal mass systems in which the lower-mass NS is
tidally wrapped around its more massive companion reach similar
temperatures, but generally tend to have more mass at lower densities
in the disk/torus \citep{oechslin:07}.

There is no unique model/EOS independent mapping $T=T(\rho)$, thus we
must explore a variety of possibilities. In
Fig.~\ref{fig:tempScripts}, we contrast our set of temperature
parametrizations with a $T(\rho)$ profile obtained from a
$1.35-1.35\,M_\odot$ simulation using the HShen EOS by
\cite{sekiguchi:11nsns} at $\sim$$12\,\mathrm{ms}$ after merger. We
consider very hot cores at 20, 30, and 40\,MeV with cold envelopes
(parametrizations c20p0, c30p0, and c40p0) and two parametrizations
with very hot cores at $30\,\mathrm{MeV}$ and cool envelopes at
$10\,\mathrm{MeV}$ and $5\,\mathrm{MeV}$, c30p10 and c30p5,
respectively. Since low-density regions have shorter neutrino cooling
times, the c30p10 and c30p5 parametrization may represent early
HMNSs, while the cold-envelope parametrizations c20p0, c30p0, and
c40p0 may correspond to late-time HMNSs. Note that the c30p10
parametrization fits the temperature profile from the
\cite{sekiguchi:11nsns} simulation quite well. Details on the
functional forms of our parametrizations can be found in
Appendix~\ref{app:temp}. For the TOV case we also consider isothermal
configurations as a limiting case.

The choice of $Y_e(\rho,T(\rho))$ is equally difficult. Before merger, the NSs
are in $\nu$-less $\beta$-equilibrium ($\mu_\nu = \mu_e + \mu_p -
\mu_n = 0 $). After merger, neutrinos are present. They are trapped in
hot dense matter ($\mu_\nu \ne 0$) and are streaming away from
low-density regions. The equilibrium $Y_e$ will shift and mixing due
to non-linear hydrodynamics in the HMNS phase will distort any initial
$Y_e(\rho,T(\rho))$ profile. 

We deem the following prescription for $Y_e$ to be the physically most
sensible: We assume that the NSNS merger occurs so rapidly that the
electron fraction $Y_e$ of the $\nu$-less $\beta$-equilibrium in the
NSs becomes the trapped postmerger lepton fraction $Y_\mathrm{lep} =
Y_e + Y_{\nu_e} - Y_{\bar{\nu}_e}$ above $\rho_\mathrm{trap}$. Using
the $\beta$-equilibrium condition with nonzero $\mu_\nu$, we solve for
$Y_e$. At densities below $\rho_\mathrm{trap}$ we transition to $Y_e$
given by $\nu$-less $\beta$-equilibrium. Details of this procedure are
given in Appendix~\ref{sec:ye}. We refer to this parametrization of
$Y_e$ as $\nu$-full $\beta$-equilibrium. In addition and for
comparison, we consider choices of constant $Y_e = 0.1$ and $Y_e$ set
according to $\nu$-less $\beta$-equilibrium. We note that
  our parameterization of $Y_e$ is ad hoc and cannot account for
  mixing and neutrino transport effects in the merger process. The
right panel of Fig.~\ref{fig:tempScripts} depicts $Y_e(\rho,T(\rho))$ as
obtained from the simulation of \cite{sekiguchi:11nsns} contrasted
with $Y_e$ profiles computed under the assumption of $\nu$-less and
$\nu$-full $\beta$-equilibrium for various temperature
parametrizations and for the $T(\rho)$ as given by the
simulation. None of the prescriptions fit the simulation-$Y_e$
particularly well, which indicates that mixing and
neutrino transport effects are important (but cannot be included
here). The $Y_e$ obtained using the temperature data from the
simulation naturally fits best, in particular at low densities where
neutrinos have decoupled from the matter and $\nu$-less
$\beta$-equilibrium holds.

In the top panels of Fig.~\ref{fig:pRatio}, we show the fractional
pressure increase due to thermal effects as a function of baryon
density for our set of temperature parametrizations for the LS220 EOS
(left panel) and the HShen EOS (right panel) as two representative
example EOS.  We also distinguish between the choices of $Y_e$
parametrization.  For the parametrizations with cold ``mantles''
(cXp0), thermal effects are most important at densities near
$\sim$$\rho_\mathrm{nuc}$ and quickly lose significance at lower and
higher densities in both EOS. The thermal pressure enhancement is at
most a factor of three (for the HShen) to five (for the LS220 EOS) for
these parametrizations. The situation is different for the cases with
hot plateaus, c30p10 and c30p5. For these, the thermal pressure is up
to $20$ times larger at low densities than predicted by the cold
EOS. The $Y_e$ parametrizations corresponding to $\nu$-full and
$\nu$-less $\beta$-equilibrium yield qualitatively and quantitatively
very similar results for both EOS.

At low densities, the $\nu$-full and $\nu$-less $\beta$-equilibrium
cases both lead to $Y_e > 0.1$ (cf.~Fig.~\ref{fig:tempScripts}). As a
consequence, the pressure in the unrealistic $Y_e = const. = 0.1$,
cXp0 parametrizations is lower than in the cold $\nu$-less case at
$\rho_b \lesssim $$10^{12.2}\,\mathrm{g}\,\mathrm{cm}^{-3}$. Due to
the logarithmic scale of Fig.~\ref{fig:pRatio}, the graphs of cXp0
with $Y_e = 0.1$ start only there and the predicted pressure
enhancement is higher than in the $\beta$-equilibrium cases, which
lead to lower $Y_e$ above
$\sim$$10^{12.2}\,\mathrm{g}\,\mathrm{cm}^{-3}$ and below
$\sim$$\rho_\mathrm{nuc}$. In the cases with hot plateau (c30p10 and
c30p5), thermal effects dominate over differences in $Y_e$ at low
densities. Finally, at $\rho > \rho_\mathrm{nuc}$, where temperature
effects are smaller, differences in $Y_e$ become important. Since the
nuclear component dominates there, lower $Y_e$ corresponds to higher
pressure (e.g., \citealt{lattimer:01}) and both $\beta$-equilibrium
cases yield $Y_e > 0.1$.

The lower panels of Fig.~\ref{fig:pRatio} depict the relative
contribution of the neutrinos to the total (hot) pressure in the HMNS
temperature and $Y_e$ parametrizations considered in this
study. While there are clear temperature (see
Eq.~\ref{eq:pressure_nu}) and $Y_e$ (through $\mu_{\nu_e}$)
dependences, neutrino pressure plays only a minor role, making up at
most $\sim$$2\%$ of the total pressure of the LS220 EOS. This is true
also for the HShen EOS with the exception of the unrealistic $Y_e =
0.1$ case in which the neutrino pressure contribution grows to
$\gtrsim$10\% of the total pressure at supranuclear densities.

Finally, we note that the temperature and $Y_e$ prescriptions
discussed here lead to regions that may be unstable to convection if
not stabilized by a positive specific angular momentum gradient (e.g.,
\citealt{tassoul:78}). The spherically and axially symmetric
equilibrium models that we construct in this study cannot account for
convection and we leave an analysis of convective instability to
future work.

\subsection{Spherically Symmetric Equilibrium Models}
\label{sec:tovmethods}
We solve the Tolman-Oppenheimer-Volkoff (TOV) equation (e.g.,
\citealt{shapteu:83}),
\begin{equation}
\label{eq:tov}
\frac{dP}{dr} = - \frac{G}{r^2}\! \left[\rho_\mathrm{b}\!\! \left(1 +
  \frac{\epsilon}{c^2} + \frac{P}{\rho_\mathrm{b} c^2}\right)\right]\!\!\! \left[ M_\mathrm{g}(r)
  + 4\pi r^3 \frac{P}{c^2} \right]\!\! \left[1 - \frac{2 G
    M_\mathrm{g}(r)}{r c^2}\right]^{-1}\!,
\end{equation}
where $r$ is the areal (circumferential) radius, $\rho_\mathrm{b}$ is
the baryon density, $\epsilon$ is the specific internal energy, and
$M_g(r)$ is the gravitational mass enclosed by radius $r$, determined
via
\begin{equation}
\frac{dM_\mathrm{g}}{dr} = 4\pi r^2 \rho_\mathrm{b} \left[1 + \frac{\epsilon}{c^2} \right]\,\,.
\end{equation}
The baryonic mass is larger and given by
\begin{equation}
\frac{dM_\mathrm{b}}{dr} = 4\pi r^2 \rho_\mathrm{b}
\left(1-\frac{2GM_\mathrm{g}(r)}{rc^2}\right)^{-1/2}\,\,.
\end{equation}

We construct the TOV solutions using a standard fourth-order
Runge-Kutta integrator on an equidistant grid with $\delta R =
10^2\,\mathrm{cm}$ zones.  After each integration sub-step, the
equation of state $P = P(\rho_\mathrm{b})$ is inverted to obtain
$\rho_\mathrm{b}$. We use a variety of $P(\rho_\mathrm{b})$
parametrizations: (\emph{i}) $T=const.$ (isothermal) with $\nu$-full
$\beta$-equilibrium above $\rho_\mathrm{trap}$ and $\nu$-less
$\beta$-equilibrium below, (\emph{ii}) $T=const.$ with $\nu$-less
$\beta$-equilibrium, (\emph{iii}) $T=const.$ with constant $Y_e =
0.1$, and (\emph{iv}) the phenomenological cXpX temperature
parametrizations with $\nu$-full $\beta$-equilibrium above
$\rho_\mathrm{trap}$ and $\nu$-less equilibrium below. We compute TOV
solutions for all EOS and define the surface of the neutron star as
the areal radius at which one of the following two conditions is true:
(\emph{i}) the pressure equals $10^{-10}$ of the central pressure;
(\emph{ii}) the pressure predicted by the integration of
Eq.~(\ref{eq:tov}) drops below the lowest pressure value available in
the equation of state table. The latter is not a limitation, because
the high-density TOV configurations considered here have steep density
and pressure profiles near their surfaces. The pressure dropping to
very small values thus indicates that the surface has been reached.

Besides the EOS, temperature, and $Y_e$ prescription, the central
baryon density $\rho_{\mathrm{b},c}$ is the only other free parameter.
Since we are interested in the maximum mass that can be supported, we
compute sequences with varying $\rho_\mathrm{b,c}$ for each EOS, but limit
ourselves to $\rho_\mathrm{nuc} < \rho_\mathrm{b,c} \le \rho_\mathrm{max,EOS}$,
where the latter is just the maximum density entry in the respective
EOS table. HMNSs with central densities below $\rho_\mathrm{nuc}$
are not realistic (cf.~\citealt{sekiguchi:11nsns}).

We make our TOV solver, all $P=P(\rho_\mathrm{b})$ tables, and the {\tt Python}
scripts used to create the results in this paper available on {\tt
  http://www.stellarcollapse.org}.

\subsection{Axisymmetric Equilibrium Models}
\label{sec:cstmethods}
We generate axisymmetric equilibrium models using the code originally
presented in \cite{cook:92} (hereafter CST; see also
\citealt{cook:94a,cook:94b}), which is based on the relativistic
self-consistent field method of Komatsu, Eriguchi \& Hachisu (1989a).
The axisymmetric equilibrium equations are solved iteratively on a
grid in $(s,\mu)$, where $s$ is a compactified radial coordinate and
$\mu = \cos{\theta}$, where $\theta$ is the usual spherical polar
angle.  Additionally, metric functions are solved using Green's
functions integrals expanded in terms of $N_l$ Legendre polynomials.
Consequently, the total numerical resolution is specified via a tuple
of $(N_s,N_\mu,N_l)$, which we set to (500, 300, 16).  The resolution
is chosen so that the resulting integral quantities of the equilibrium
solution (e.g., its gravitational mass) are precise to about one part
in $10^3$. The surface of the star is defined by an enthalpy contour
which is specified in the code by setting a surface energy density.
This energy density has a default value of $7.9$ g cm$^{-3}$, and we
have checked that increasing its value by a factor of $10^6$ leaves
the physical quantities of the solution unchanged to our stated
general error level of $10^{-3}$.

An axisymmetric HMNS equilibrium configuration is constructed by the
CST code based on choices of (i) a barotropic EOS, (ii) a rotation
law, (iii) the rotation rate, and, (iv) the maximum mass-energy
density $E_\mathrm{max} = [\rho_b(1+\epsilon/c^2)]_\mathrm{max}$ of
the configuration.

In order to keep the size of the parameter space manageable, we
restrict rotating configurations to the LS220 and HShen EOS and set up
barotropic versions using the temperature and composition
parametrizations described in \S\ref{sec:tempcomp}. Since the EOS
obtained with $\nu$-full and $\nu$-less $\beta$-equilibrium differ
only very mildly (cf. Fig.~\ref{fig:pRatio}), we construct rotating
configurations under the simple assumption of $\nu$-less
$\beta$-equilibrium.

We employ the `$j-const.$' rotation law (see, e.g., CST), which is
commonly used in the literature for HMNS models (e.g.,
\citealt{baumgarte:00}).  The degree of differential rotation is
parametrized by $\tilde{A}$ \footnote{Note that $\tilde{A} =
  1/\hat{A}$, where $\hat{A}$ is the same $\hat{A}$ as used in
  \cite{baumgarte:00}.}.  In the Newtonian limit, this rotation law
becomes $\Omega = \Omega_c / (1 + \tilde{A}^2 \varpi^2/r_e^2)$, where
$\varpi$ is the cylindrical radius, $r_e$ is the radius of the star at
its equator, and $\Omega_c$ is the central angular velocity. For
$\tilde{A} = 0$, one recovers uniform rotation, while for large
$\tilde{A}$, the specific angular momentum becomes constant (i.e.,
$\Omega \propto \varpi^{-2}$ in the Newtonian limit).  We explore
values of $\tilde{A}$ between $0$ and $1$. The latter value of
$\tilde{A}$ corresponds to roughly a factor of two decrease of the
angular velocity from the center to the HMNS surface, which is in the
ball park of what is found in merger simulations (e.g.,
\citealt{shibata:2005ss}). Once the rotation law is fixed, the
rotation rate is determined by specifying the \textit{axis ratio}
$r_\mathrm{p/e}$, defined as the ratio of the HMNS radius along the
pole $r_\mathrm{p}$ divided by the radius at the equator
$r_\mathrm{e}$.

The final parameter to be chosen is the maximum energy density of the
configuration. For simplicity and consistency with the choice of
variables for the TOV solutions discussed in \S\ref{sec:tovmethods},
we set $E_\mathrm{max}$ by choosing a maximum baryon density
$\rho_\mathrm{b,{max}}$ and obtain $E(\rho_{\mathrm{b,max}})$ from the
EOS.

For each choice of EOS, $\rho_\mathrm{b,\mathrm{max}}$, and
$\tilde{A}$, we compute a sequence of models with increasing rotation
rate, stepping down from $r_\mathrm{p/e} = 1$ (the
nonrotating TOV case) until we reach mass shedding or until the code
fails to converge to an equilibrium solution. In the case of uniform
rotation ($\tilde{A} = 0$) the sequence always ends at mass shedding,
the resulting rotating neutron star has spheroidal shape, and the
maximum and central density coincide ($\rho_{b,\mathrm{max}} =
\rho_\mathrm{c}$). Differentially rotating sequences, on the other
hand, can bifurcate into two branches: one with
$\rho_{b,\mathrm{max}} = \rho_\mathrm{c}$ and spheroidal geometry and
one with an off-center location of $\rho_{b,\mathrm{max}}$ and
quasitoroidal shape. For differentially rotating models, the CST
solver generally fails to converge to a solution at
$r_\mathrm{p/e}$  before mass shedding and, therefore,
possibly before the maximum mass for a given configuration is
reached. This limitation means that the maximum masses we state for
differentially rotating models are to be interpreted as lower bounds
on the true maximum masses.  The code developed by \cite{ansorg:03} is
far more robust than CST for such extreme configurations and these
authors have argued that with increasing degree of differential
rotation, arbitrarily large masses could be supported in extremely
extended tori, but such configurations are unlikely to be
astrophysically relevant.

\section{Results: Spherically Symmetric Models}
\label{sec:TOVresults}

Our main interest is in how temperatures in the range encountered in
HMNS of NSNS postmerger simulations change the maximum mass that can
be supported.  Since baryonic mass is a conserved quantity and can be
related to the number of baryons present in the individual NSNS before
merger (modulo a small amount of potential ejecta), we treat it as a
the most important variable and define the maximum gravitational
masses $M_\mathrm{g}^\mathrm{max}$ as the gravitational mass at which
$M_\mathrm{b}^\mathrm{max}$ is maximal.  We consider the isothermal
TOV solution as a limiting case of maximal thermal support but note
that such configurations with $T\gtrsim 5-8\,\mathrm{MeV}$ develop
very large, non-degenerate envelopes at the low end of the central
baryon densities $\rho_\mathrm{b,c}$ considered here. With increasing
temperature, degeneracy is more and more lifted at those densities and
the TOV model approaches an isothermal sphere whose pressure is
dominated by relativistic non-degenerate pairs and whose mass and
radius become infinite. We discard such solutions.

\begin{figure}[t]
\centering
\includegraphics[width=\columnwidth]{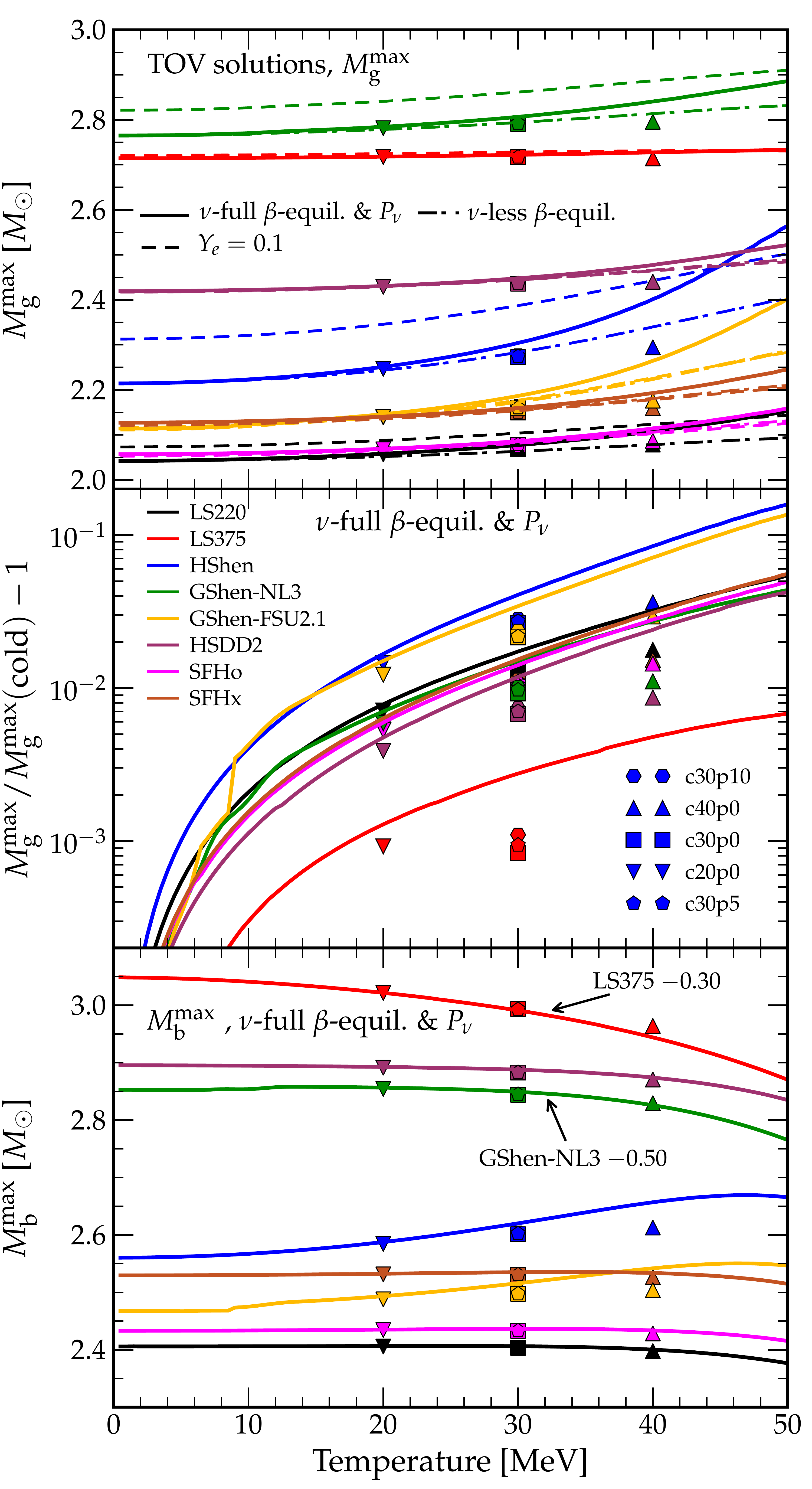}
\caption{Effect of temperature $T$ on the maximum masses of TOV
  solutions.  {\bf Top panel}: gravitational mass
  $M_\mathrm{g}^\mathrm{max}$ at the maximum baryonic mass for
  $T=const.$ configurations (lines) and parametrized cXpX profiles
  (symbols). The cXpX solutions are computed only for $\nu$-full
  $\beta$-equilibrium. With increasing $T$,
  $M_\mathrm{g}^\mathrm{max}$ increases. This trend is independent of
  $Y_e$ prescription, but the sensitivity to $Y_e$ is highly EOS
  dependent.  {\bf Center panel}: Relative increase of
  $M_\mathrm{g}^\mathrm{max}$ with $T$ for solutions in $\nu$-full
  $\beta$-equilibrium. The increase is modest and below
  $\sim10\%$ even in the $T=const.$ case.  {\bf Bottom
    panel:} Maximum baryonic mass $M_\mathrm{b}^\mathrm{max}$ that can
  be supported as a function of temperature. For most EOS, there is
  little variation in $M_\mathrm{b}^\mathrm{max}$ at low $T$, but the
  increasing thermal contribution to the TOV energy density
  (cf.~Eq.~\ref{eq:tov}) leads to a \textit{decrease} of
  $M_\mathrm{b}^\mathrm{max}$ for $high$-T solutions.  A linear
  vertical shift of -0.30 (-0.50) $M_\odot$ has been applied to the
  LS375 (GShen-NL3) curves to enhance the vertical dynamic range of
  the plot.  } \label{fig:tov}
\end{figure}

The results of our TOV calculations are summarized by
Fig.~\ref{fig:tov} for all considered EOS. We provide numerical
results in Tab.~\ref{tab:tov} for fiducial isothermal cold
($T=0.5\,\mathrm{MeV}$) and parametrized temperature choices.

In the top panel of Fig.~\ref{fig:tov}, we show the maximum
gravitational mass (defined as $M_\mathrm{g}$ at
$M_\mathrm{b}^\mathrm{max}$) as a function of isothermal
temperature for our three $Y_e$ prescriptions. The considered EOS show
a great degree of variation in their sensitivity to $Y_e$
prescriptions, but the overall trend is clear: increasing temperature
generally leads to increasing $M_\mathrm{g}^\mathrm{max}$. The
fractional increase over the cold value, however, is not large, as
shown by the center panel. The HShen and GShen-FSU2.1 RMF TOV stars
are the most sensitive to temperature variations\footnote{See, e.g.,
  \citealt{hempel:12} for a discussion of EOS physics and temperature
  dependence of various EOS models.}, but even their maximum
gravitational TOV mass increases only by $\sim$$12-15\%$ at isothermal
$T=50\,\mathrm{MeV}$. The cXpX temperature parametrizations, shown
as symbols in Fig.~\ref{fig:tov} located at their respective central
temperatures, generally follow the trend of the isothermal sequences
for each EOS, but their $M_\mathrm{g}^\mathrm{max}$ enhancement is
systematically lower, since they are only centrally hot.

\begin{deluxetable}{llcccc}
\tablecolumns{6} 
\tablewidth{0pc} 

\tablecaption{Summary of TOV Results for all EOS. \label{tab:tov}}

\tablehead{EOS &
  T($\rho$) & $M^\mathrm{max}_\mathrm{b}$ &
  $M^\mathrm{max}_\mathrm{g}$ & $R$ & $\rho_c$ \\ & & ($M_\odot$)
  & ($M_\odot$) & (km) & ($10^{15}\,\mathrm{g\,cm}^{-3}$)}

\startdata
\\ [-2ex]
LS220, $\nu$-less & 0.5\,MeV & 2.406 & 2.042 & 10.63 & 1.863\\
LS220, $\nu$-full &c20p0 &2.434 &2.068 &10.69 &1.873\\
&c30p0 &2.433 &2.078 &10.89 &1.840\\
&c30p10 &2.433 &2.079 &11.86 &1.840\\
&c30p5 &2.433 &2.078 &11.23 &1.840\\
&c40p0 &2.428 &2.087 &11.07 &1.808\\ 
\hline \\ [-1.5ex]
LS375, $\nu$-less  & 0.5\,MeV &3.349 &2.715 &12.34 &1.243\\
LS375, $\nu$-full  
&c20p0 &3.322 &2.717 &12.59 &1.232\\
&c30p0 &3.294 &2.717 &12.68 &1.221\\
&c30p10 &3.293 &2.718 &13.49 &1.221\\
&c30p5 &3.293 &2.717 &12.95 &1.221\\
&c40p0 &3.264 &2.714 &12.75 &1.210\\
\hline \\ [-1.5ex]
HShen, $\nu$-less  
& 0.5\,MeV &2.560 &2.214 &12.59 &1.357\\
HShen, $\nu$-full 
&c20p0 &2.584 &2.246 &13.17 &1.321\\
&c30p0 &2.601 &2.273 &13.48 &1.276\\
&c30p10 &2.604 &2.277 &15.08 &1.276\\
&c30p5 &2.603 &2.275 &14.01 &1.276\\
&c40p0 &2.613 &2.295 &13.69 &1.243\\
\hline \\ [-1.5ex]
GShen-NL3, $\nu$-less 
&0.5\,MeV&3.353 &2.765 &13.34 &1.115\\
GShen-NL3, $\nu$-full 
&c20p0 &3.354 &2.781 &13.51 &1.098\\
&c30p0 &3.344 &2.791 &13.70 &1.081\\
&c30p10 &3.346 &2.793 &15.04 &1.081\\
&c30p5 &3.345 &2.792 &14.30 &1.081\\
&c40p0 &3.330 &2.796 &13.86 &1.070\\
\hline \\ [-1.5ex]
GShen-FSU2.1, $\nu$-less 
&0.5\,MeV&2.468 &2.114 &11.67 &1.505\\
GShen-FSU2.1, $\nu$-full 
&c20p0 &2.488 &2.140 &12.15 &1.474\\
&c30p0 &2.497 &2.159 &12.40 &1.428\\
&c30p10 &2.502 &2.164 &14.30 &1.420\\
&c30p5 &2.497 &2.160 &12.44 &1.428\\
&c40p0 &2.504 &2.176 &12.56 &1.398\\
\hline \\ [-1.5ex]
HSDD2, $\nu$-less 
&0.5\,MeV&2.896 &2.419 &11.92 &1.395\\
HSDD2, $\nu$-full 
&c20p0 &2.891 &2.429 &12.28 &1.381\\
&c30p0 &2.883 &2.436 &12.43 &1.367\\
&c30p10 &2.884 &2.437 &13.47 &1.367\\
&c30p5 &2.883 &2.436 &12.79 &1.367\\
&c40p0 &2.871 &2.440 &12.55 &1.353\\
\hline \\ [-1.5ex]
SFHo, $\nu$-less 
&0.5\,MeV&2.433 &2.057 &10.31 &1.906\\
SFHo, $\nu$-full 
&c20p0 &2.434 &2.068 &10.67 &1.884\\
&c30p0 &2.433 &2.078 &10.86 &1.862\\
&c30p10 &2.434 &2.079 &11.81 &1.862\\
&c30p5 &2.433 &2.078 &11.21 &1.851\\
&c40p0 &2.428 &2.087 &11.03 &1.829\\
\hline \\ [-1.5ex]
SFHx, $\nu$-less 
&0.5\,MeV&2.529 &2.127 &10.79 &1.722\\
SFHx, $\nu$-full 
&c20p0 &2.531 &2.139 &11.18 &1.705\\
&c30p0 &2.530 &2.150 &11.37 &1.688\\
&c30p10 &2.531 &2.151 &12.39 &1.688\\
&c30p5 &2.531 &2.150 &11.72 &1.688\\
&c40p0 &2.527 &2.160 &11.51 &1.671
\enddata
\tablecomments{``$\nu$-less''
  indicates neutrino-less $\beta$-equilibrium, which we use only for
  the ``cold'' configurations. ``$\nu$-full'' indicates neutrino-full
  $\beta$-equilibrium with neutrino pressure. $T(\rho)$ is the
  temperature parametrization, $M_\mathrm{b}^\mathrm{max}$ is the
  maximum baryonic mass, $M_\mathrm{g}^\mathrm{max}$ is the
  gravitational mass at the maximum baryonic mass, $R$ is the radius
  of the $M_\mathrm{b}^\mathrm{max}$ configuration, and $\rho_c$ is
  the central baryon density at which $M_\mathrm{b}^\mathrm{max}$
  obtains.}
\end{deluxetable}

The lower panel of Fig.~\ref{fig:tov} depicts the change of the
maximum baryonic TOV mass $M_\mathrm{b}^\mathrm{max}$ with increasing
temperature. For most EOS, $M_\mathrm{b}^\mathrm{max}$ stays roughly
constant at low temperatures, but \textit{decreases} at high
temperatures. This shows that the increase in
$M_\mathrm{g}^\mathrm{max}$ in the TOV solutions is primarily due to
thermal contributions to the total mass-energy density. Since it is
the mass-energy density, and not just the baryonic mass, which sources
curvature (the relativistic gravitational field), the thermal effects
lead to a decrease in $M_\mathrm{b}^\mathrm{max}$ with temperature
even if $M_\mathrm{g}^\mathrm{max}$ is still increasing. The HShen and
GShen-FSU2.1 are the only two EOS that exhibit an increase of
$M_\mathrm{b}^\mathrm{max}$ at intermediate to high temperatures, but
they too reverse this trend at isothermal $T \gtrsim
50\,\mathrm{MeV}$. The LS375 EOS, on the other hand, has monotonically
decreasing $M_\mathrm{b}^\mathrm{max}$ with $T$, which was seen before
by \cite{oconnor:11}. The more realistic cXpX temperature
parametrizations show a similar trend as their isothermal
counterparts, but for the HShen and GShen-FSU2.1 EOS, the increase in
$M_\mathrm{b}^\mathrm{max}$ at intermediate $T$ is smaller in these
only centrally-hot parametrized models.

It is interesting to compare our findings with the results of
\cite{oconnor:11}, who studied black hole formation through
protoneutron star collapse in failing core-collapse supernovae.  These
authors found much larger maximum baryonic and gravitational masses of
their protoneutron stars at the onset of collapse than reported here.
The collapsing protoneutron stars in their study have moderately-high
central temperatures $T \lesssim 40\,\mathrm{MeV}$. However, at $\rho
\approx 4\times 10^{14} - 10^{15}\,\mathrm{g\,cm}^{-3}$, a region of
extremely hot material with $T\gtrsim 80-100\,\mathrm{MeV}$ is present
due to compression of multiple $M_\odot$ of accreted shock-heated
material. \cite{oconnor:11} demonstrated that this extremely hot
region is responsible for the observed thermal enhancement of the
maximum protoneutron star mass. In NSNS mergers the situation is quite
different and fully dynamical NSNS merger simulations have not found
such extremely hot high-density regions
\citep[e.g.,][]{sekiguchi:11nsns,oechslin:07}. It is thus unlikely
that the findings of \cite{oconnor:11} apply to the merger HMNS case.

\begin{figure*}[t!]
\centering
\includegraphics[width=\columnwidth]{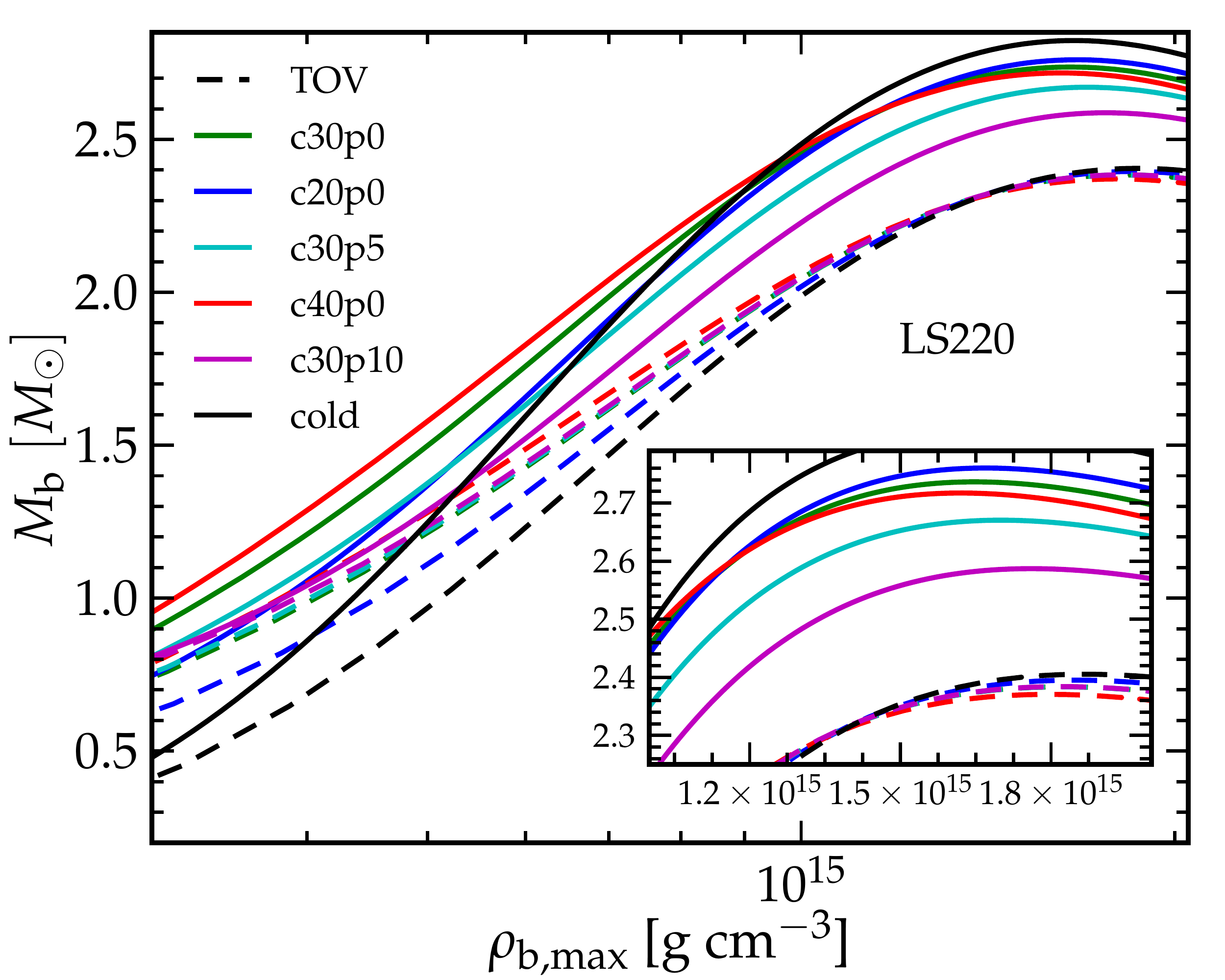}
\hspace{0.05\columnwidth}
\includegraphics[width=\columnwidth]{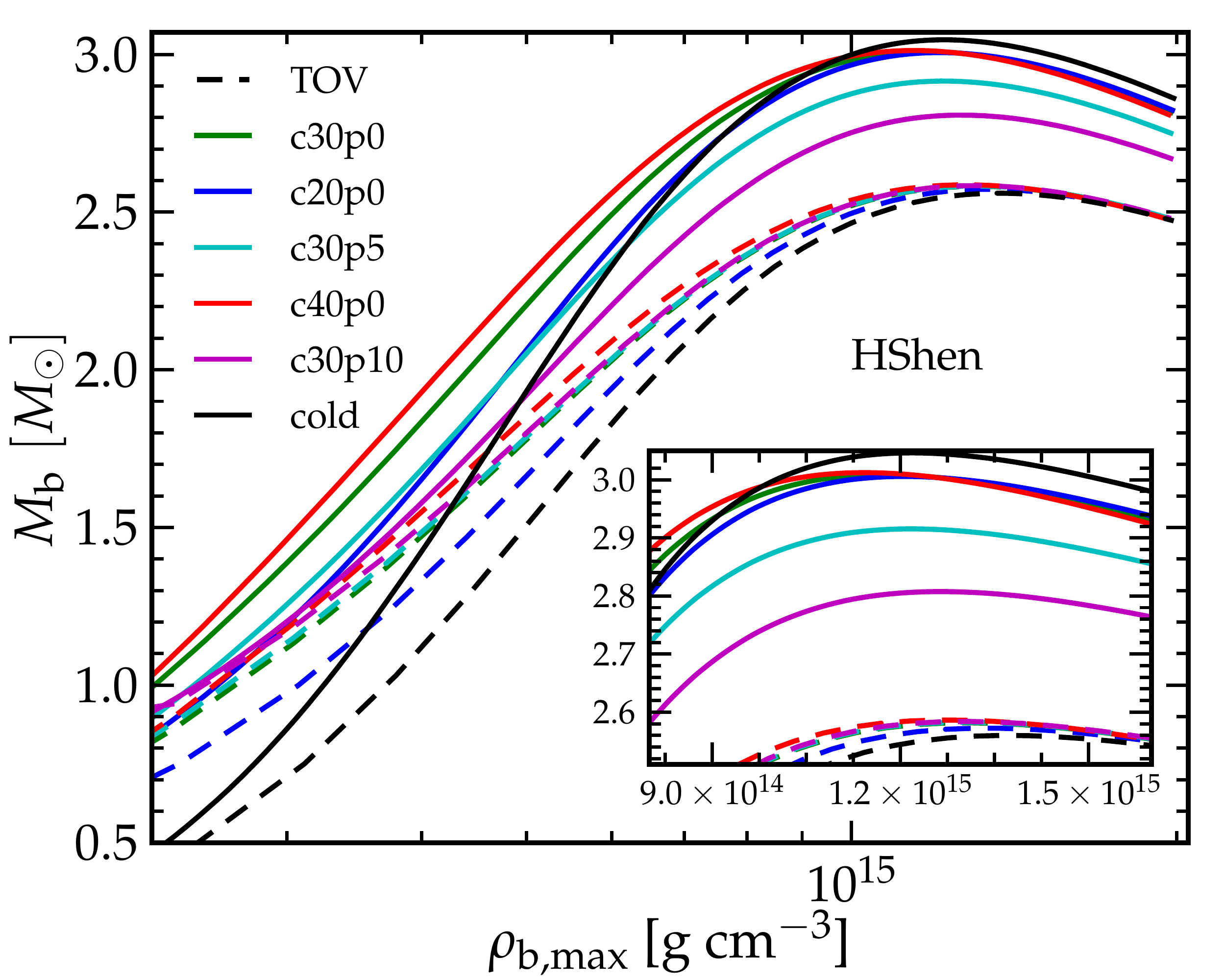}
\caption{Baryonic mass $M_\mathrm{b}$ as a function of maximum baryon
  density $\rho_\mathrm{b,max}$ of uniformly rotating ($\tilde{A}=0$)
  equilibrium models at the mass-shedding limit for different
  temperature prescriptions (solid lines). We also plot the
  corresponding TOV sequences (dashed lines) and show results for the
  the LS220 EOS (\textbf{left panel}) and HShen EOS (\textbf{right
    panel}). There is a large thermal enhancement of $M_\mathrm{b}$ at
  low densities, but the sequences converge towards the cold
  supramassive limit as the maximum density increases and the
  configurations become more compact.  }
\label{fig:mbUniformRot}
\end{figure*}

\section{Results: Axisymmetric Models in Rotational Equilibrium}
\label{sec:CSTresults}

\subsection{Uniformly Rotating Configurations}
\label{sec:results_2Duni}

It has been widely recognized that uniform rotation can support a
supramassive neutron star against gravitational collapse (see,
e.g.~\citealt{friedman:86,friedman:87}).  A supramassive neutron star
is defined as a stable neutron star with a mass greater than the
maximum mass of a TOV star with the same EOS (CST).  At a given
central density, the mass that may be supported rises with increasing
angular velocity until the material on the NS's equator becomes
unbound (the mass-shedding limit).  This leads to the
\emph{supramassive limit}, a well defined maximum mass for uniformly
rotating NSs with a specified EOS.

\begin{figure*}
\centering
\includegraphics[width=\columnwidth]{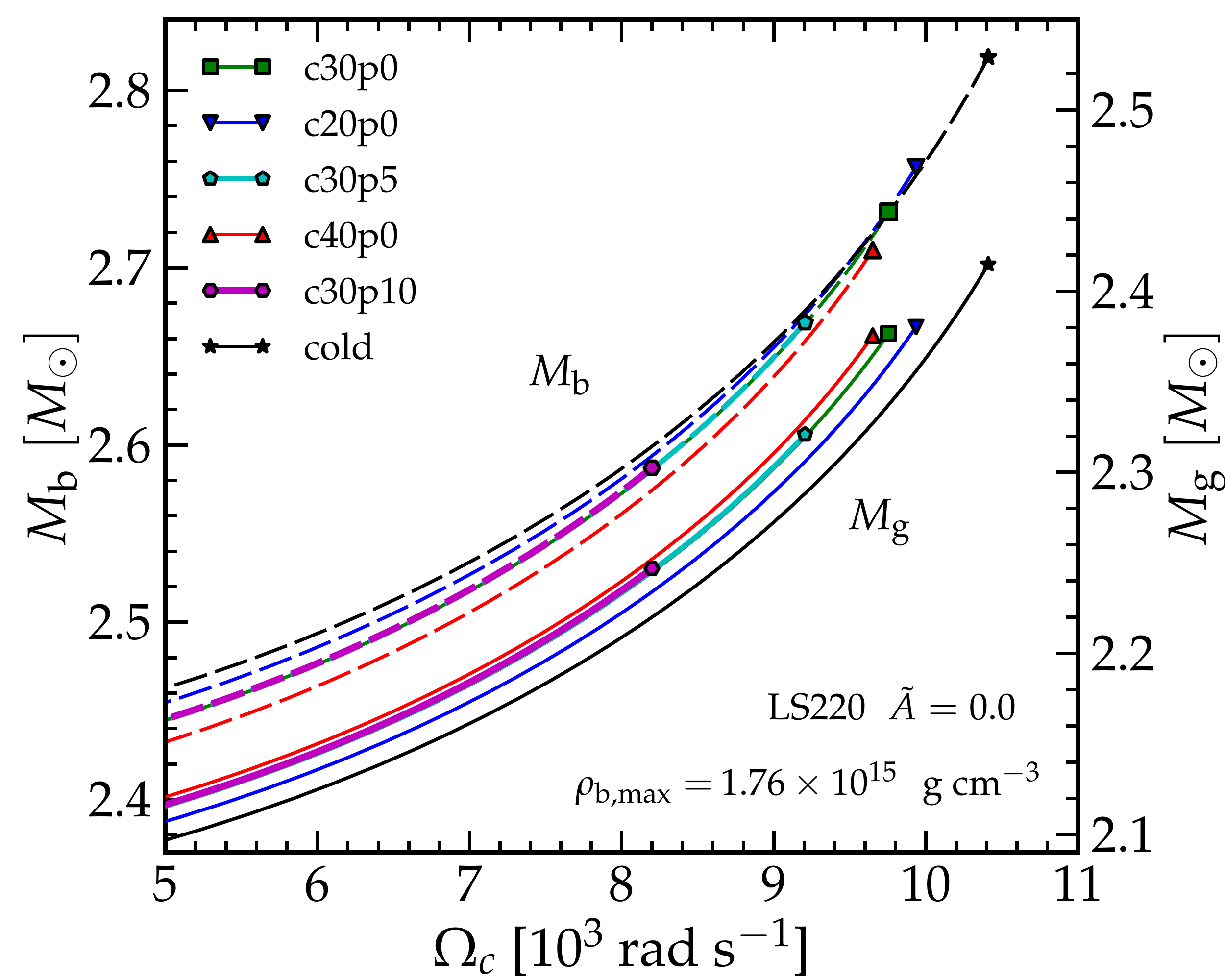}
\hspace{0.05\columnwidth}
\includegraphics[width=\columnwidth]{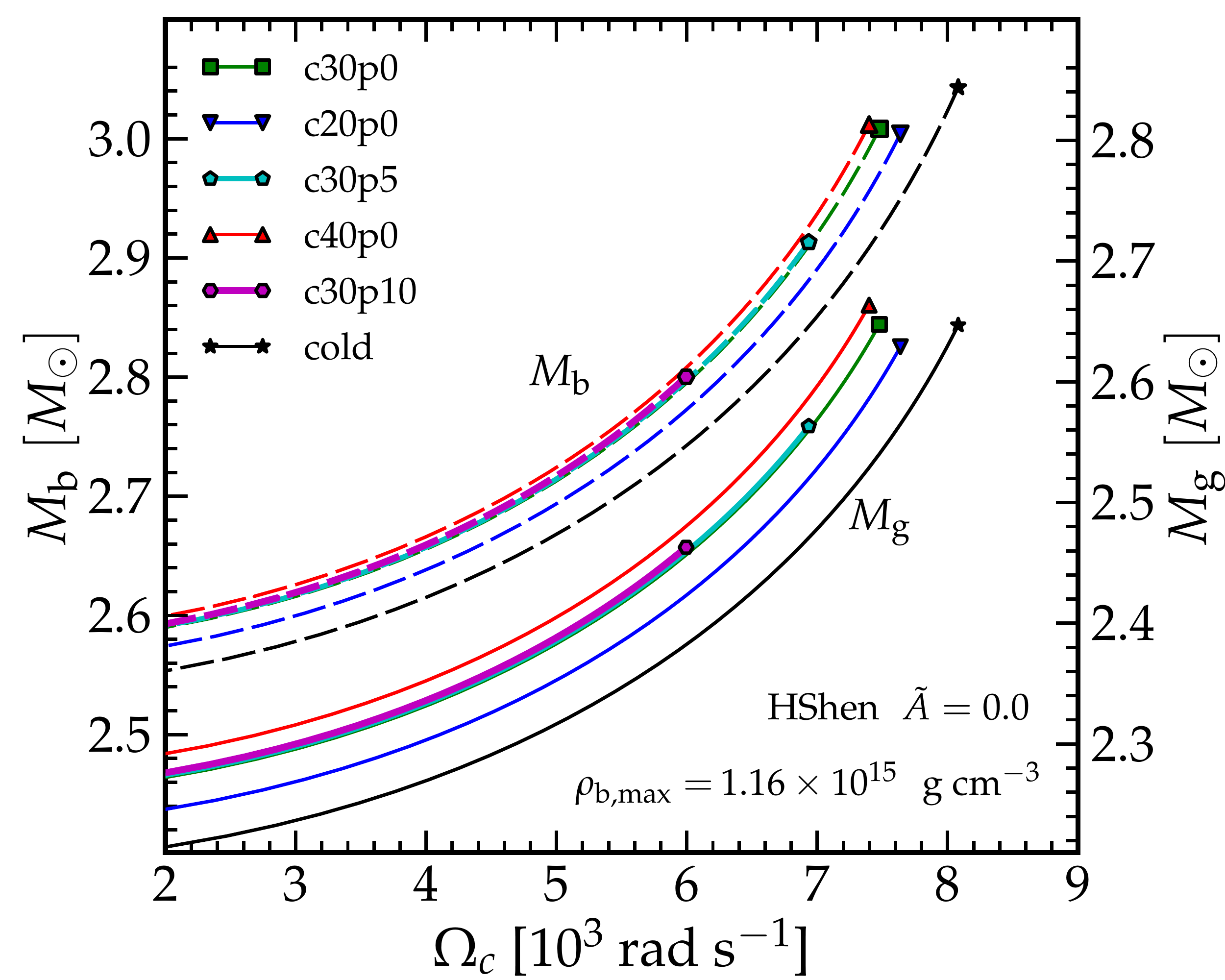}
\caption{The gravitational mass ($M_\mathrm{g}$, solid lines, right
  ordinate) and baryonic mass ($M_\mathrm{b}$, dashed lines, left
  ordinate) as a function of angular velocity $\Omega$ for uniformly
  spinning models at a fixed density near the density that yields the
  maximum $M_\mathrm{b}$ for the LS220 (left panel) and the HShen EOS
  (right panel). The sequences terminate at the mass-shedding limit,
  which is the point with the maximum angular velocity for a specific
  temperature prescription. Configurations with higher temperatures
  and, in particular, the c30p5 and c30p10 models with
  high-temperature plateaus at low densities, have larger radii than
  colder models and thus reach the mass-shedding limit at lower
  angular velocities. Hence, such models have lower maximum masses at
  the supramassive limit than colder models. Note that hotter models
  with the LS220 have lower baryonic masses than colder models.}
\label{fig:omegaUniformRot}
\end{figure*}

In Fig.~\ref{fig:mbUniformRot}, we plot the baryonic mass
$M_\mathrm{b}$ as a function of maximum baryon density for TOV and
uniformly rotating mass-shedding sequences obtained with the LS220 EOS
(left panel) and the HShen EOS (right panel). Focusing first on the
TOV sequences, one notes that at low central densities
($\rho_\mathrm{b} \lesssim \,\mathrm{few}\,\times \rho_\mathrm{nuc}$),
$M_\mathrm{b}$ is significantly increased by thermal effects. This is
because the mean density $\bar{\rho}_b$ of such configurations is in
the regime in which thermal pressure is of greatest relevance
(cf.~Fig.~\ref{fig:pRatio}) and can alter the structure of the bulk of
the NS.  This carries over to the uniformly rotating case. The
extended hot configurations reach mass shedding at lower angular
velocities than their cold counterparts, but the extended, low
$\bar{\rho}_b$ cores of hot configurations receive sufficient
rotational support to yield a higher $M_\mathrm{b}$. This, however,
is the case only for centrally-hot cXp0 configurations. Models with
hot envelopes (with parametrizations c30p5 and c30p10) benefit
less from rotational support.

With increasing maximum density, the baryonic masses of the TOV models
for different temperature parametrizations converge for a given
EOS. Near the density at which the maximum mass is reached, the
increase in $M_\mathrm{b}$ in hot configurations has turned into a
slight decrease for models computed with the LS220 EOS and has dropped
to $\lesssim$$5\%$ for the HShen EOS (see also
Fig.~\ref{fig:tov}). The mass-shedding sequences show a more complex
behavior with increasing maximum density. As in the TOV case, the mean
density $\bar{\rho}_b$ of the NSs increases and less material is
experiencing enhanced pressure support due to high temperatures in the
cXp0 models. Hence, these models move towards the
$M_\mathrm{b}^\mathrm{max}$ of the cold supramassive limit (see the
inset plots in Fig.~\ref{fig:mbUniformRot}). For both EOS, the
$M_\mathrm{b}^\mathrm{max}$ of hot configurations are all lower than
the cold value. The cXp0 models reach supramassive limits that are
within less than $2\%$ of the cold supramassive limit for both
EOS. The c30p10 and c30p5 models, on the other hand, have
$M_\mathrm{b}^\mathrm{max}$ that are $\sim$$5-10\%$ lower than the
cold supramassive limit for both EOS. Table~\ref{tab:uniform}
summarizes key parameters of the hot and cold configurations at the
supramassive limit.

\begin{deluxetable*}{lccccccc}
  \tablecolumns{5} \tablewidth{0pc} 
\tablecaption{Uniformly Rotating Neutron Stars at the Supramassive Limit} 
\tablehead{ 
Model 
&$\rho_\mathrm{b,max}$ 
&$M_\mathrm{b}^\mathrm{max}$  
&$M_\mathrm{g}^\mathrm{max}$
&$r_e$
&$r_{p/e}$
&$\Omega$
&$T/|W|$ \\
&($10^{15}\,\mathrm{g}\,\mathrm{cm}^{-3}$)
&($M_\odot$) 
&($M_\odot$) 
&(km)
&
&($10^3\,\mathrm{rad}\,\mathrm{s}^{-1}$)
&
}
\startdata 
LS220 cold    & 1.653 & 2.823 & 2.419 & 14.429 & 0.566 & 10.096 & 0.118\\
LS220 c20p0    & 1.652 & 2.760 & 2.384 & 14.788 & 0.574 & \phantom{0}9.647 & 0.106\\
LS220 c30p0    & 1.652 & 2.737 & 2.382 & 15.000 & 0.576 & \phantom{0}9.441 & 0.103\\
LS220 c30p5    & 1.710 & 2.671 & 2.322 & 15.300 & 0.587 & \phantom{0}9.031 & 0.088\\
LS220 c30p10    & 1.769 & 2.587 & 2.247 & 16.130 & 0.599 & \phantom{0}8.215 & 0.066\\
LS220 c40p0    & 1.625 & 2.717 & 2.383 & 15.201 & 0.577 & \phantom{0}9.262 & 0.101\\
\hline\\ [-1.5ex]
HShen cold    & 1.220 & 3.046 & 2.649 & 17.101 & 0.564 & 8.233 & 0.117\\
HShen c20p0    & 1.196 & 3.006 & 2.629 & 17.760 & 0.573 & 7.745 & 0.105\\
HShen c30p0    & 1.171 & 3.009 & 2.648 & 18.173 & 0.574 & 7.511 & 0.103\\
HShen c30p5    & 1.228 & 2.916 & 2.564 & 18.665 & 0.588 & 7.086 & 0.084\\
HShen c30p10    & 1.261 & 2.808 & 2.467 & 20.070 & 0.604 & 6.238 & 0.060\\
HShen c40p0    & 1.139 & 3.012 & 2.664 & 18.474 & 0.574 & 7.355 & 0.101
\enddata
\tablecomments{Summary of mass-shedding
    uniformly rotating supramassive neutron star configurations at the
    maximum mass for each EOS and temperature prescription. These
    models are in $\nu$-less $\beta$-equilibrium (see
    \S\ref{sec:tempcomp}). $\rho_\mathrm{b,max}$ is the central
    density of the model with the maximum baryonic mass
    $M_\mathrm{b}^\mathrm{max}$. $M_\mathrm{g}^\mathrm{max}$ is the
    gravitational mass at the $\rho_\mathrm{b,max}$ at which
    $M_\mathrm{b}^\mathrm{max}$ occurs. $r_e$ is the equatorial radius, $r_{p/e}$ 
    is the axis ratio, $\Omega$ is the angular velocity, and $T/|W|$
    is the ratio of rotating kinetic energy $T$ to gravitational energy $|W|$.
}
\label{tab:uniform}
\end{deluxetable*}

The systematics of the supramassive limit with temperature
prescription becomes clear when considering
Fig.~\ref{fig:omegaUniformRot}. This figure shows the baryonic mass
$M_\mathrm{b}$ and gravitational mass $M_\mathrm{g}$ for uniformly
rotating NSs as a function of angular velocity $\Omega$ for the LS220
and HShen EOS at fixed densities near the maximum of
$M_\mathrm{b}(\rho_\mathrm{b,max})$ (see Table~\ref{tab:uniform}). At
fixed angular velocity below mass shedding, hotter configurations
always yield higher $M_\mathrm{g}$ than their colder counterparts. For
the LS220 EOS, as in the TOV case discussed in the previous section
\ref{sec:TOVresults}, hotter configurations have lower
$M_\mathrm{b}$. In the case of the HShen EOS, which generally yields
less compact equilibrium models, the opposite is true, but the
increase in $M_\mathrm{b}$ caused by thermal support is smaller than
the increase in $M_\mathrm{g}$.

With increasing $\Omega$, the mass-shedding limit is approached and
hotter configurations systematically reach the mass shedding limit at
lower angular velocities. The reason for this is best illustrated by
comparing c30p0 models with c30p10 and c30p5 models, which have a
high-temperature plateau at low densities of $10\,\mathrm{MeV}$ and
$5\,\mathrm{MeV}$, respectively. At low angular velocities, all c30pX
models show the same thermal increase in $M_\mathrm{g}$. However, the high
pressure at low densities in the c30p10 and c30p5 models leads to
significantly larger radii compared to the model without temperature
plateau. Consequently, as $\Omega$ is increased, the configurations
with plateau reach the mass-shedding limit at lower $\Omega$. For the
LS220 EOS, the c30p10 sequence terminates at
$\sim$$8200\,\mathrm{rad}\,\mathrm{s}^{-1}$, the c30p5 sequence
terminates at $\sim$$9200\,\mathrm{rad}\,\mathrm{s}^{-1}$, while the
c30p0 sequence does not terminate before
$\sim$$9800\,\mathrm{rad}\,\mathrm{s}^{-1}$. The HShen model sequences
show the same qualitative trends.

\subsection{Differentially Rotating Configurations}
\label{sec:results_2Ddiff}

Differential rotation can provide centrifugal support at small radii
while allowing a NS configuration to stay below the mass-shedding
limit at its equatorial surface. Differentially rotating equilibrium
configurations have been shown to support masses well in excess of the
supramassive limit
\citep[e.g.,][]{ostriker:66,baumgarte:00,morrison:04}. Such
configurations are referred to as ``hypermassive''.  However, since
there is (mathematically speaking) an infinite number of possible
differential rotation laws, it is impossible to define a formal
``hypermassive limit'' for the maximum mass of HMNSs in the way it is
possible for uniformly rotating supramassive NSs. Nevertheless, we can
study the systematics of the supported baryonic (and gravitational)
masses with variations in the HMNS temperature profile, maximum baryon
density, and degree and rate of differential rotation for the rotation
law considered in this study, which is not drastically different from
what is found in merger simulations (e.g., \citealt{shibata:2005ss}).

\begin{figure*}[t]
\centering
\includegraphics[width=\columnwidth]{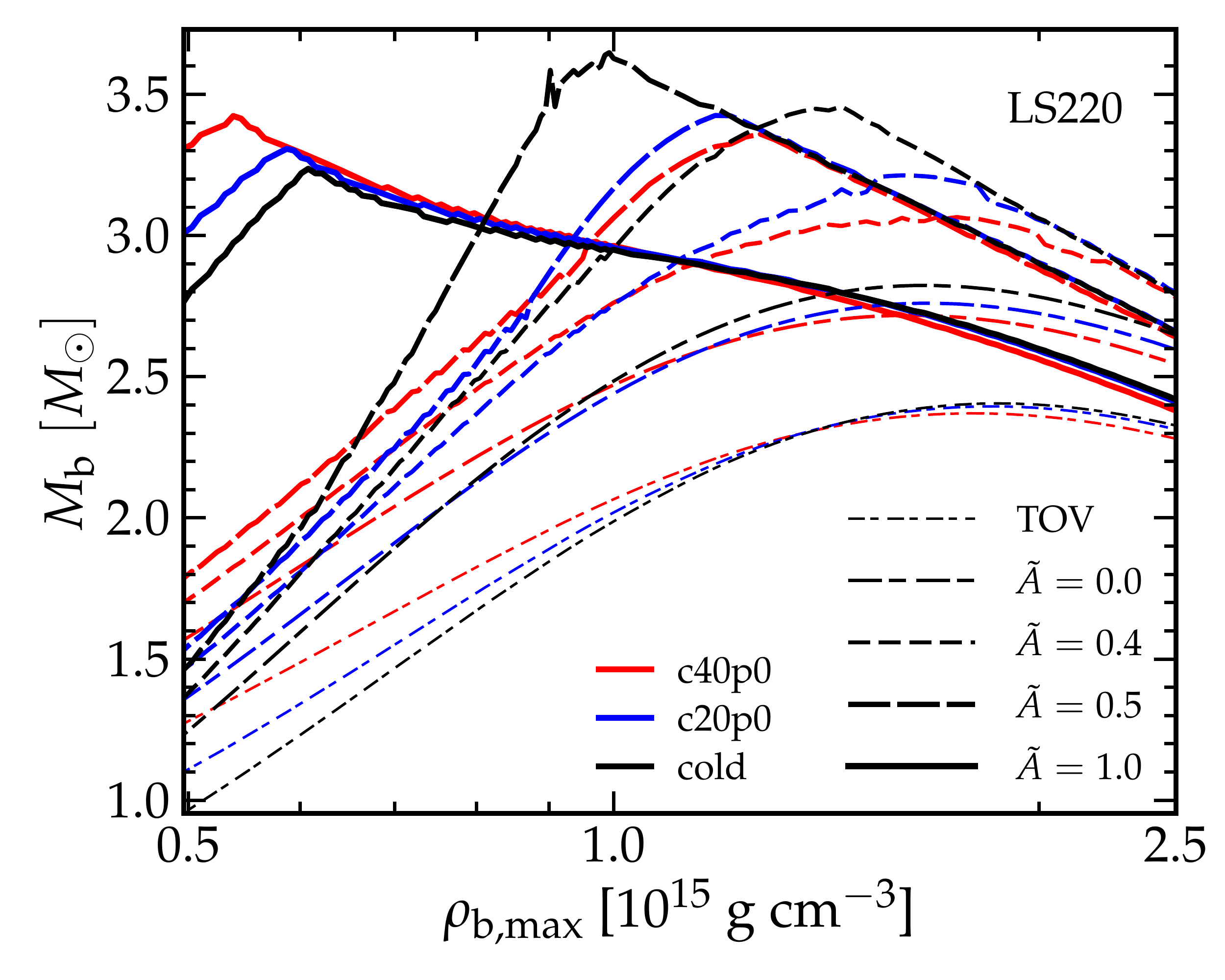}
\includegraphics[width=\columnwidth]{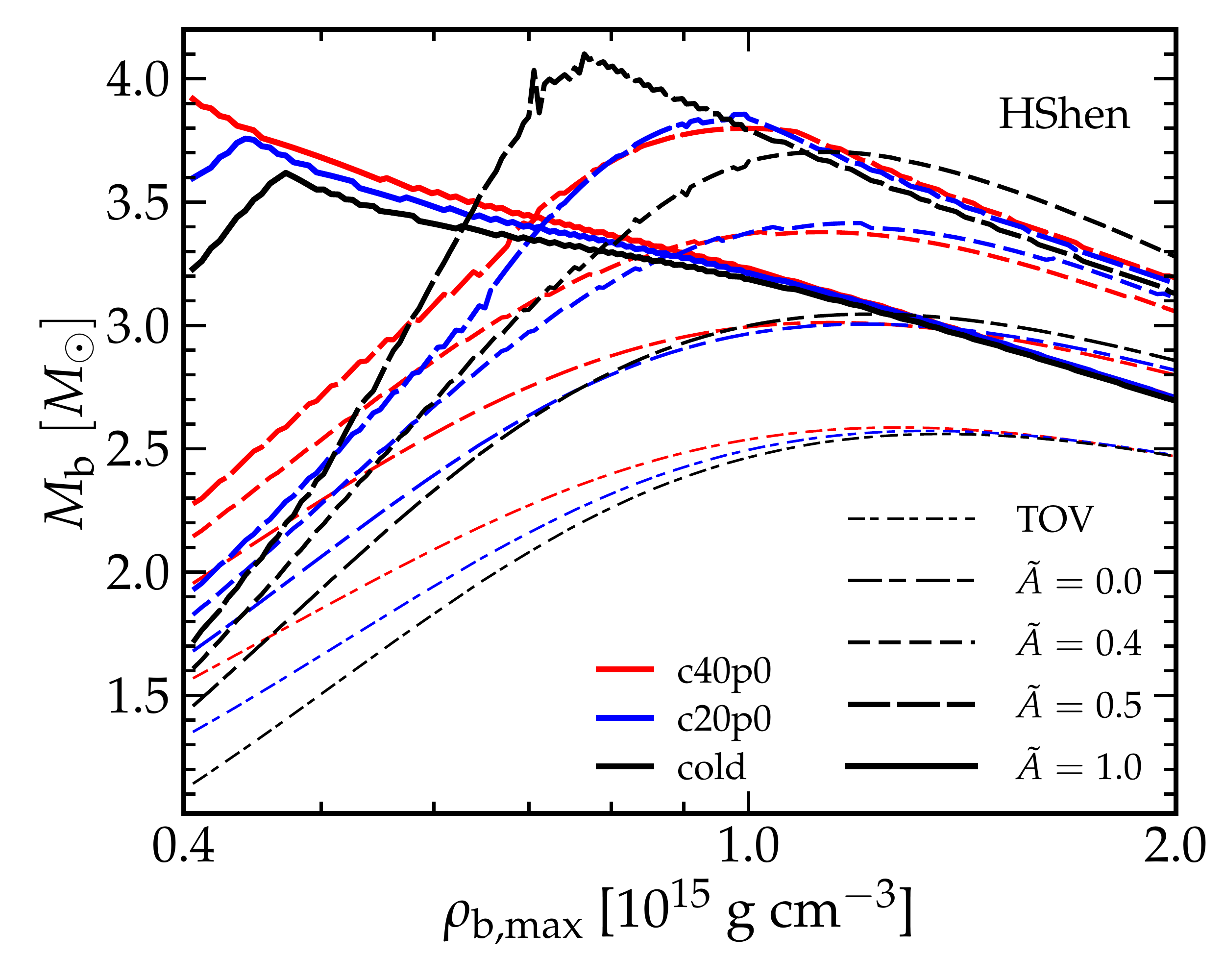}
\caption{ Maximum baryonic mass configurations for sequences of
  uniformly rotating ($\tilde{A}=0$) and differentially rotating
  ($\tilde{A} = \{0.4,0.5,1.0\}$) models with cold, c20p0, and c40p0
  temperature parametrizations and the LS220 EOS (left panel) and
  HShen EOS (right panel). We note that for differentially rotating
  models these curves represent lower limits on the maximum baryonic
  mass (i.e., the solver fails to converge at lower axis
    ratios without reaching the true mass shedding limit).  We limit the
  sequences to models with $T/|W| \lesssim 0.25$ and this limit
  defines the rising part of the graphs for $\tilde{A}=1$ at low
  densities.  We show the TOV case (thinnest and shortest dash-dotted
  lines) for comparison. The raggedness of the curves with $\tilde{A}
  \gtrsim 0.4$ is a consequence of finite resolution in the parameter
  $r_{p/e}$ that is varied to find the maximum mass at a given
  $\rho_\mathrm{b,max}$. Thermal effects are most pronounced at low
  densities and for high $\tilde{A}$. For uniform and moderate
  differential rotation, hotter models have lower global maximum
  $M_\mathrm{b}$ than colder models.}
\label{fig:mbDiffRot1}
\end{figure*}

In Fig.~\ref{fig:mbDiffRot1}, we show the supported baryonic mass
$M_\mathrm{b}$ as a function of maximum baryon density
$\rho_\mathrm{b,max}$ for cold, c20p0, and c40p0 temperature
prescriptions, both EOS, and for different choices of $\tilde{A}$.
The curves represent configurations with the minimum $r_\mathrm{p/e}$
at which an equilibrium solution is found by the CST solver (i.e., the
most rapidly spinning setup). Note that the peaks of these curves
represent only \emph{lower} limits on the maximum HMNS mass. In
addition, we plot only solutions with ratios $T/|W|$ of rotational
kinetic energy $T$ to gravitational energy $|W|$ below $25\%$, since
more rapidly spinning models would be dynamically nonaxisymmetrically
unstable \citep{chandrasekhar69c,baiotti:07}. It is this limit which
defines the rising branch of the $M_\mathrm{b}(\rho_\mathrm{b,max})$
curve at the lowest densities in Fig.~\ref{fig:mbDiffRot1} for
$\tilde{A} = 1.0$. Note that many of these configurations may still be
unstable to secular rotational instabilities or rotational shear
instabilities (e.g., \citealt{watts:05,ott:07prl,corvino:10}).

The overall shape of the $M_\mathrm{b}(\rho_\mathrm{b,max})$ curves in
Fig.~\ref{fig:mbDiffRot1} is qualitatively similar to what is shown in
Fig.~1 of \cite{baumgarte:00} for $\Gamma = 2$ polytropes and Fig.~2
of \cite{morrison:04} for the cold FPS EOS \citep{friedman:81}. The
LS220 and HShen EOS yield qualitatively very similar results, but the
supported HMNS masses found by the CST solver are, as expected,
systematically higher for models with the HShen EOS than for those
using the LS220 EOS. One notes, however, interesting variations with
temperature prescription.  At low $\rho_\mathrm{b,max}$, thermal
pressure support leads to increased $M_\mathrm{b}$ and more
differentially rotating configurations have higher
$M_\mathrm{b}$. Sequences with $\tilde{A} \lesssim 0.5$ show similar
systematics with density and temperature prescription as the uniformly
spinning ones discussed in \S\ref{sec:results_2Duni}: As the density
increases, hot configurations converge towards the cold sequence and
reach their maximum $M_\mathrm{b}$ near and below the maximum of the
cold sequence. Models with $\tilde{A} \gtrsim 0.5$, on the other hand,
have more steeply rising curves with $\rho_\mathrm{b,max}$ and are
discontinuous (i.e., exhibit a ``kink'') at their global maxima. At
these points quasitoroidal solutions appear. Furthermore, the slope of
the curve describing (as a function of $\rho_{\mathrm{b,max}}$) the
axis ratios $r_\mathrm{p/e}$ at which the solver stops converging
discontinuously changes sign. We attribute this behavior, which was
also observed by \citealt{morrison:04}, to a bifurcation of the
sequence between models, which continue shrinking in axis ratio until
they become completely toroidal ($r_\mathrm{p/e} = 0$), and less
extreme models that stay quasitoroidal or spheroidal . Beyond the
``kink'' in $\tilde{A}\gtrsim 0.5$ sequences, thermal effects play
little role.

\begin{figure*}[t]
\centering
\includegraphics[width=\columnwidth]{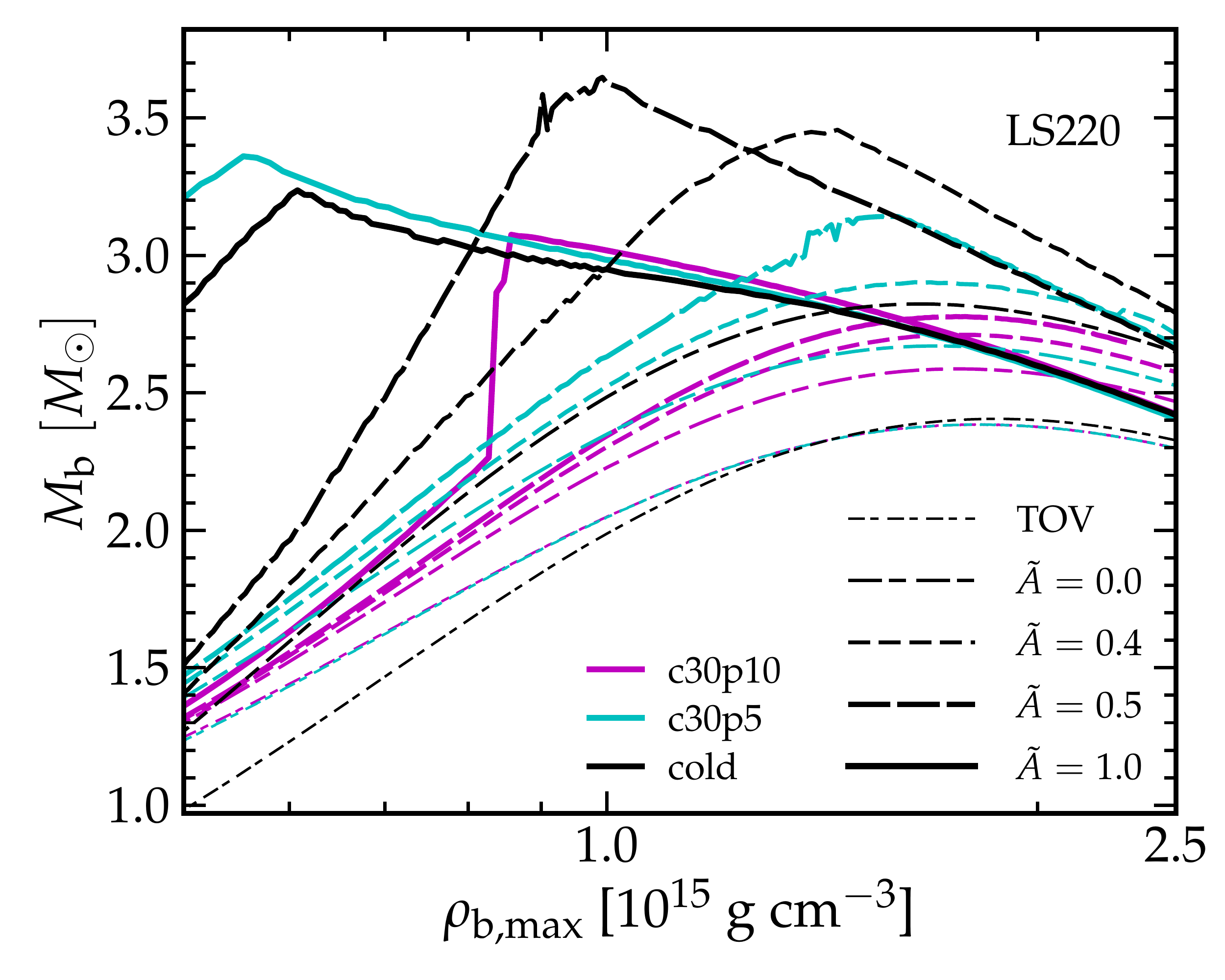}
\includegraphics[width=\columnwidth]{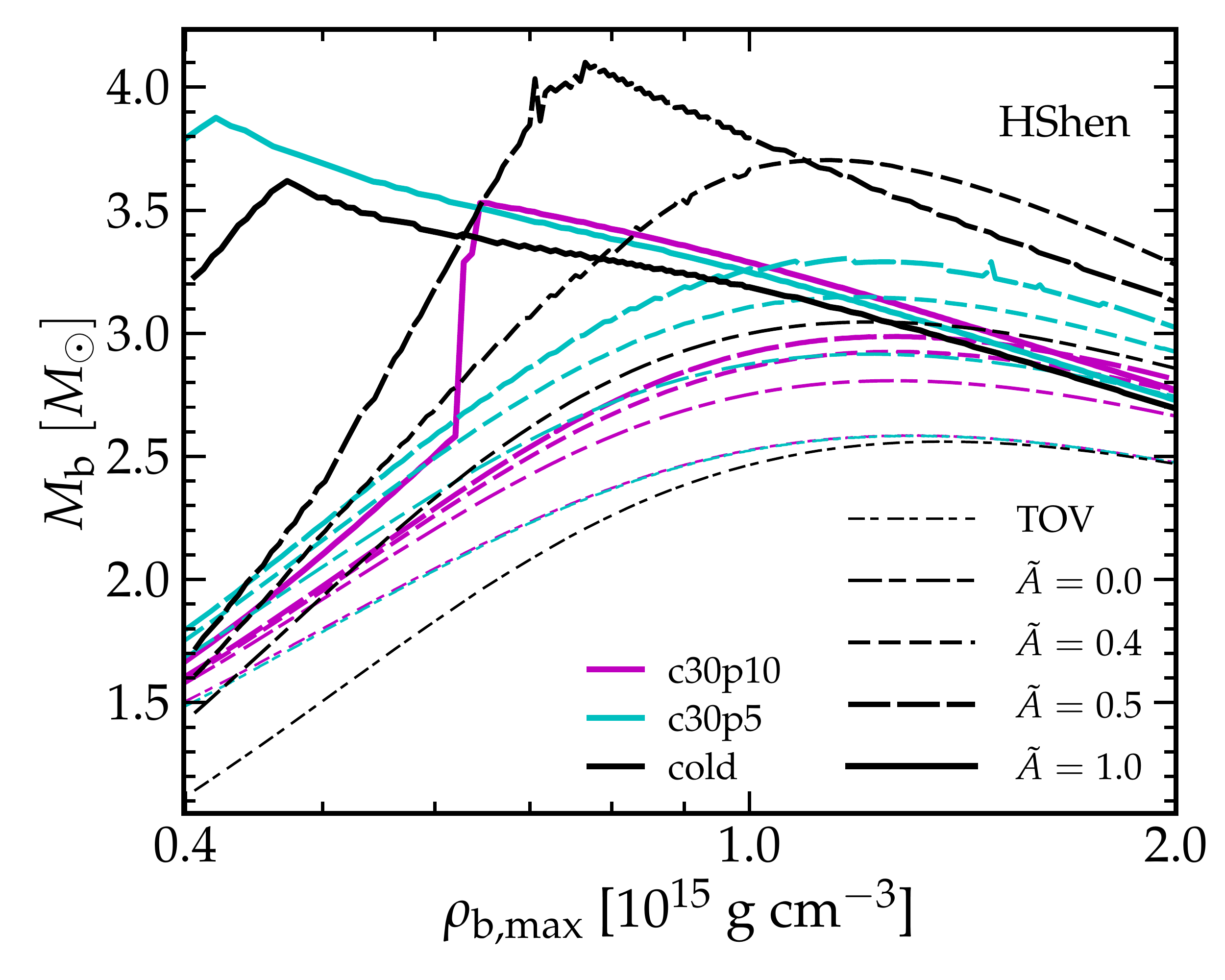}
\caption{Same as Fig.~\ref{fig:mbDiffRot1}, but comparing cold
  configurations with models with the c30p5 and c30p10 temperature
  prescriptions, which have a hot plateau at low densities.  The
  overall systematics are the same for the LS220 EOS (left panel) and
  the HShen EOS (right panel). In the TOV case, $M_\mathrm{b}$ is
  thermally enhanced at low densities, but the global maximum of
  $M_\mathrm{b}$ of hot configurations is near that of the cold TOV
  solution. Uniformly and moderately differentially rotating sequences
  of c30p10 and c30p5 models have systematically smaller maximum
  masses than cold models throughout the considered density
  range. Only very differentially rotating models ($\tilde{A}\gtrsim
  0.7$; $\tilde{A}=1.0$ shown here) exhibit a thermal enhancement of
  the maximum mass at low to intermediate densities. The c30p10
  sequence for $\tilde{A}=1.0$ exhibits a discontinuous jump, which
  occurs when the sequence transitions from spheroidal to
  quasitoroidal shape. See text for discussion.  }
\label{fig:mbDiffRot2}
\end{figure*}

The lower bounds of the range of $\rho_\mathrm{b,max}$ shown in the
two panels of Fig.~\ref{fig:mbDiffRot1} (and also
Fig.~\ref{fig:mbDiffRot2}) are chosen for the following reason: Fully
dynamical merger simulations by, e.g, \cite{sekiguchi:11nsns,
  baiotti:08a,shibata:2005ss,kiuchi:2009jt,bauswein:12,thierfelder:11},
all suggest a rule of thumb that the postmerger maximum baryon density
of the HMNS is typically not less than $\sim$$80\%$ of the central
density of the progenitor NSs. We can derive a rather solid
EOS-dependent lower limit on $\rho_\mathrm{b,max}$ for HMNS remnants
from (equal mass) NSNS mergers in the following way: In order to form
a HMNS, constituent equal-mass NSs must at the very least have a mass
that is 50\% of the maximum mass in the cold TOV limit. Hence, the
premerger central density must at least be that of a TOV solution
with $M_\mathrm{b} = 0.5 M_\mathrm{b}^\mathrm{max,TOV}$. Using the
aforementioned empirical result from merger simulations, we arrive at
\begin{equation}
\rho_\mathrm{b,min} = 0.8 \rho_\mathrm{b,TOV}( M_\mathrm{b} =
M_\mathrm{b,max}/2)\,\,.
\label{eq:lowerrho}
\end{equation}
For the LS220 EOS, $\rho_\mathrm{b,TOV}( M_\mathrm{b} =
M_\mathrm{b,max}/2) \sim 5.8\times 10^{14}
\,\mathrm{g}\,\mathrm{cm}^{-3}$ and occurs at $M_\mathrm{b}$
($M_\mathrm{g}$) of $1.19\,M_\odot$ ($1.10\,M_\odot$). For the HShen
EOS, $\rho_\mathrm{b,TOV}( M_\mathrm{b} = M_\mathrm{b,max}/2) \sim
4.4\times 10^{14}\,\mathrm{g}\,\mathrm{cm}^{-3}$ and occurs at
$M_\mathrm{b}$ ($M_\mathrm{g}$) of $1.28\,M_\odot$ ($1.20\,M_\odot$).
Applying the density cut given by Eq.~(\ref{eq:lowerrho}) excludes
most dynamically nonaxisymmetrically unstable configurations.

Figure~\ref{fig:mbDiffRot2}, like Fig.~\ref{fig:mbDiffRot1}, shows
baryonic mass as a function of maximum baryon density for both EOS and
a variety of $\tilde{A}$, but contrasts models c30p5 and c30p10, which
have hot plateaus at low densities, with cold models. The qualitative
features discussed in the following are identical for both EOS. In the
TOV case and at low densities, $M_\mathrm{b}$ is enhanced primarily by
the hot core, since nonrotating solutions are compact and
dominated by $\rho_\mathrm{b} \gtrsim 10^{14}\,\mathrm{g\,cm}^{-3}$,
where the high-temperature plateaus play no role. At higher densities,
the $M_\mathrm{b}$ curves of hot models converge to near or below the
cold TOV maximum $M_\mathrm{b}$. The situation is different for
uniformly and moderately differentially rotating models ($\tilde{A}
\lesssim 0.5$). Rotation shifts these configurations to lower mean
densities and the hot plateaus lead to equatorially bloated
solutions. These reach their minimum $r_\mathrm{p/e}$ for which a
solution can be found at lower angular velocities. Hence, centrifugal
support is weaker and the configuration with the hottest plateau has
the lowest $M_\mathrm{b,max}$. The behavior is different at high
degrees of differential rotation ($\tilde{A} = 1$). The cold and the
c30p5 models are HMNSs and quasitoroidal already at the lowest
densities shown in Fig.~\ref{fig:mbDiffRot2}. The c30p5 sequence has
slightly larger $M_\mathrm{b}$ than the cold sequence. The c30p10
sequence, however, is spheroidal at low $\rho_\mathrm{b,max}$ and then
discontinuously transitions to the quasitoroidal branch, which is
marked by a large jump in $M_\mathrm{b}$.

\begin{deluxetable*}{lcccccccc}
  \tablecolumns{5} \tablewidth{0pc} 
\tablecaption{Differentially Rotating Hypermassive Neutron Stars} 
\tablehead{ 
Model 
&$\rho_\mathrm{b,max}$ 
&$M_\mathrm{b}^\mathrm{max}$  
&$M_\mathrm{g}^\mathrm{max}$
&$r_e$
&$r_{p/e}$
&$\tilde{A}$
&$\Omega_c$
&$T/|W|$ \\
&($10^{15}\,\mathrm{g}\,\mathrm{cm}^{-3}$)
&($M_\odot$) 
&($M_\odot$) 
&(km)
&
&
&($10^3\,\mathrm{rad}\,\mathrm{s}^{-1}$)
&
}
\startdata 
LS220 cold    & 0.993 & 3.648 & 3.140 & 17.258 & 0.376 & 0.5 & 15.476 & 0.244\\
LS220 c20p0    & 0.852 & 3.573 & 3.124 & 18.538 & 0.364 & 0.6 & 15.047 & 0.243\\
LS220 c30p0    & 0.706 & 3.568 & 3.167 & 19.611 & 0.344 & 0.7 & 14.888 & 0.249\\
LS220 c30p5    & 0.600 & 3.413 & 3.064 & 21.870 & 0.320 & 0.9 & 14.461 & 0.250\\
LS220 c30p10    & 0.990 & 3.090 & 2.723 & 19.208 & 0.421 & 0.9 & 16.330 & 0.187\\
LS220 c40p0    & 0.692 & 3.597 & 3.211 & 19.931 & 0.344 & 0.7 & 14.677 & 0.249\\
\hline\\ [-1.5ex]
HShen cold    & 0.766 & 4.101 & 3.562 & 19.800 & 0.372 & 0.5 & 13.450 & 0.245\\
HShen c20p0    & 0.641 & 4.076 & 3.585 & 21.352 & 0.360 & 0.6 & 13.042 & 0.245\\
HShen c30p0    & 0.532 & 4.099 & 3.650 & 22.305 & 0.344 & 0.7 & 13.131 & 0.249\\
HShen c30p5    & 0.517 & 3.942 & 3.527 & 24.371 & 0.340 & 0.8 & 12.426 & 0.243\\
HShen c30p10    & 0.646 & 3.529 & 3.141 & 23.521 & 0.400 & 1.0 & 13.934 & 0.196\\
HShen c40p0    & 0.514 & 4.148 & 3.708 & 22.701 & 0.344 & 0.7 & 12.888 & 0.249
\enddata
\tablecomments{ Summary of the differentially rotating HMNS
  configurations with the largest baryonic masses for each EOS and
  temperature prescription. These configurations are obtained in a
  sequence from $\tilde{A} = 0$ to $\tilde{A} = 1$ with spacing
  $\delta\tilde{A} = 0.1$ and are to be seen as lower bounds on the
  maximum achievable masses. The sequences considered here exclude
  dynamically nonaxisymmetrically unstable models with ratio of
  rotational kinetic energy to gravitational energy $T/|W| >
  0.25$. The quantities listed in the table are the following:
  $\rho_\mathrm{b,max}$ is the baryon density at which the maximum
  baryonic mass $M_\mathrm{b}^\mathrm{max}$ occurs,
  $M_\mathrm{g}^\mathrm{max}$ is the gravitational mass at that
  density, $r_e$ is the equatorial radius of the configuration,
  $r_{p/e}$ is its axis ratio, $\tilde{A}$ is the differential
  rotation parameter at which $M_\mathrm{b}^\mathrm{max}$
  obtains. $\Omega_c$ is the central angular velocity of the
  configuration and $T/|W|$ is its ratio of rotational kinetic energy
  to gravitational energy. We note that the accuracy of the results
  listed in this table is set by the step size in $r_{p/e}$, which
  we set to $\delta r_{p/e} = 0.004$.
}
\label{tab:differential}
\end{deluxetable*}

\begin{figure}[t]
\centering
\includegraphics[width=\columnwidth]{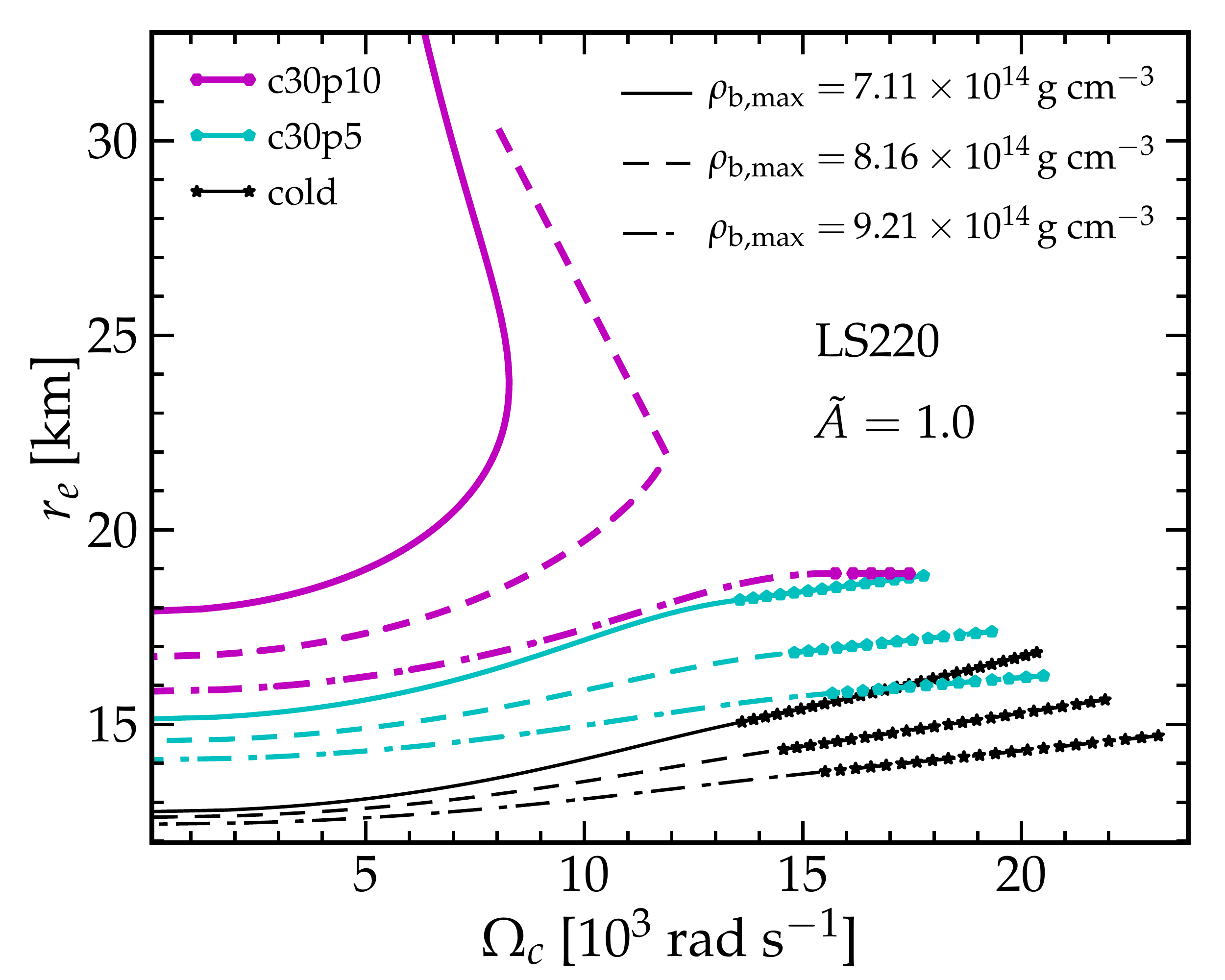}
\caption{ Equatorial radii $r_e$ vs.\ central angular velocity
  $\Omega_c$ in sequences parametrized by the axis ratio $r_{p/e}$
  for models using the LS220 EOS, differential rotation parameter
  $\tilde{A}=1.0$, and cold, c30p5, and c30p10 temperature
  parametrizations. We show curves for three densities, two below the
  discontinuous jump of the c30p10 curve in Fig.~\ref{fig:mbDiffRot2}
  and one above. At the same density, hotter configurations have
  larger radii and transition to quasitoroidal shape (marked by dots)
  at higher $\Omega_c$. The transition between spheroidal and
  quasitoroidal shape is discontinuous in $\rho_\mathrm{b,max}$ for
  critical models at the minimum $r_{p/e}$ that can be found (shown in
  Figs.~\ref{fig:mbDiffRot1} and \ref{fig:mbDiffRot2}), but smooth in
  $r_{p/e}$ at fixed $\rho_\mathrm{b,max}$. The low-density sequences
  with the c30p10 temperature prescription ($10$-$\mathrm{MeV}$ plateau
  at low densities; see \S\ref{sec:tempcomp}) become double valued in
  $\Omega_c$ with increasing $r_\mathrm{p/e}$, stay spheroidal and have
  very large $r_e$.
}
\label{fig:reVsOmegaDiffRot}
\end{figure}

\begin{figure}[t]
\centering
\includegraphics[width=\columnwidth]{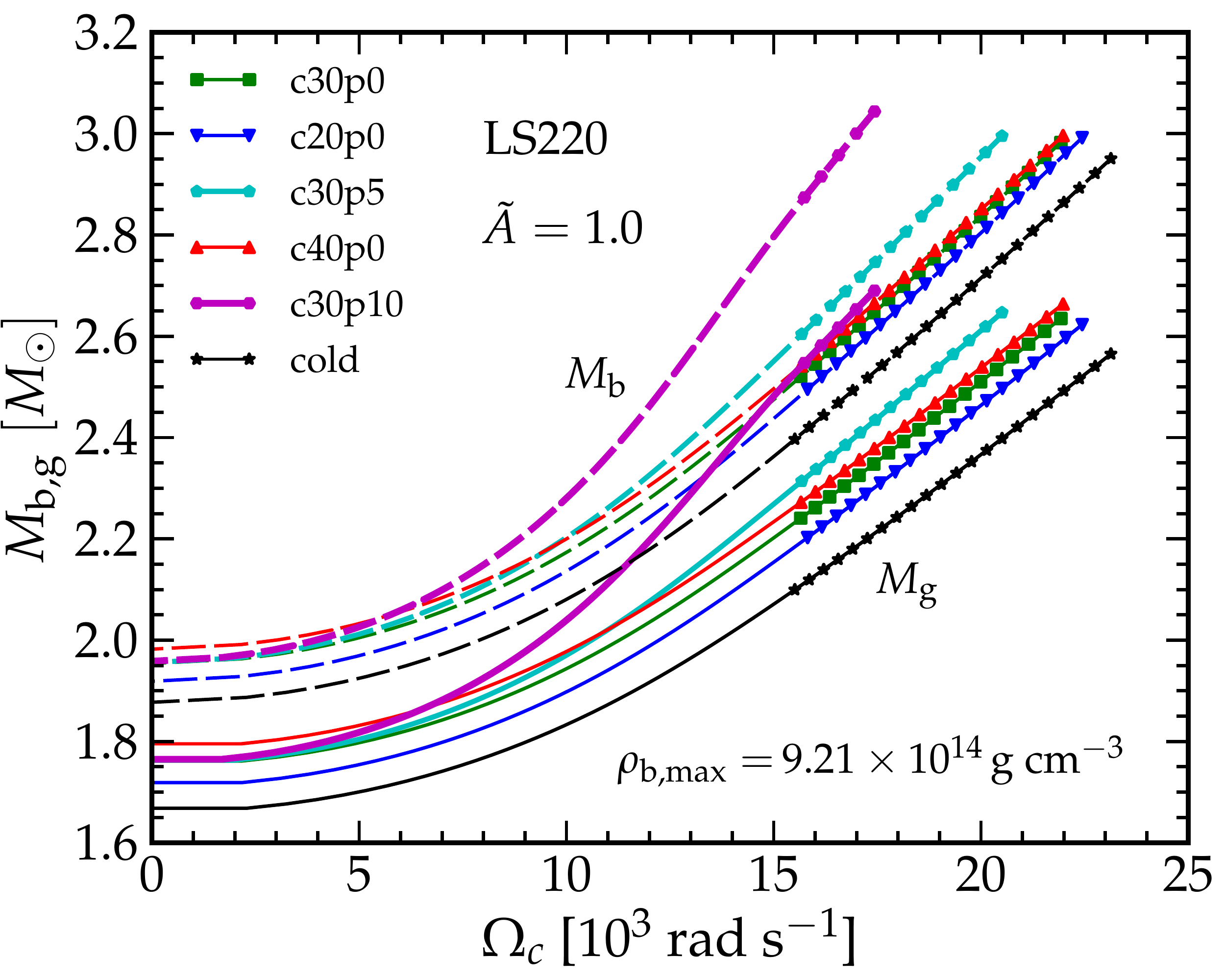}
\caption{Baryonic mass $M_\mathrm{b}$ and gravitational mass
  $M_\mathrm{g}$ vs. central angular velocity $\Omega_c$ parametrized
  by the axis ratio $r_\mathrm{p/e}$ at fixed degree of differential
  rotation $\tilde{A}=1$, and fixed maximum density of
  $\rho_\mathrm{b,max} = 9.21\times
  10^{14}\,\mathrm{g\,cm}^{-3}$. Curves for all temperature
  parametrizations are shown for the LS220 EOS. Quasitoroidal
  configurations are marked by symbols and the transitions between
  spheroidal and quasitoroidal solutions are smooth. The end points of
  all graphs correspond to the values plotted in
  Figs.~\ref{fig:mbDiffRot1} and \ref{fig:mbDiffRot2} for the various
  temperature prescriptions at $\tilde{A}=1.0$ and the
  $\rho_\mathrm{b,max}$ chosen here. Sequences with hot plateaus
  (using temperature prescriptions c30p5 and c30p10) exhibit
  significant thermal enhancements of $M_\mathrm{b}$ and
  $M_\mathrm{g}$ at rapid rotation rates, but have lower maximum
  rotation rates due to their larger radii.}
\label{fig:mVsOmegaDiffRot}
\end{figure}

In order to illustrate this discontinuous behavior further, we plot in
Fig.~\ref{fig:reVsOmegaDiffRot} the equatorial radius of equilibrium
solutions as a function of central angular velocity at $\tilde{A} = 1$
and for three different fixed $\rho_\mathrm{b,max}$. We show curves
obtained with the LS220 EOS for the cold, c30p5, and c30p10
temperature prescriptions. The curves are parametrized by decreasing
$r_\mathrm{p/e}$ and terminate at the smallest value at which the
solver converges. The three densities are chosen so that the first two
are below and the third is above the jump of the c30p10 curve in
Fig.~\ref{fig:mbDiffRot2}. At all $\rho_\mathrm{b,max}$, the hot
configurations have significantly larger radii than the cold models,
but decreasing $r_\mathrm{p/e}$ leads to increasing $\Omega_c$ and
only modest radius changes for cold and c30p5 models.  This is very
different for the c30p10 sequence.  At $\rho_\mathrm{b,max} =
7.11\times 10^{14}\,\mathrm{g\,cm}^{-3}$ these models do not become
quasitoroidal and the $r_e-\Omega_c$ mapping becomes double-valued as
the decrease in $r_\mathrm{p/e}$ turns from a decrease of
$r_\mathrm{p}$ at nearly fixed $r_\mathrm{e}$ and increasing
$\Omega_c$ into a steep increase of $r_\mathrm{e}$ and a decrease of
$\Omega_c$. As $\rho_\mathrm{b,max}$ increases, less material is at
low densities where thermal pressure support is strong in the c30p10
models. Consequently, the solutions are more compact and stay so to
smaller $r_\mathrm{p/e}$. $\rho_\mathrm{b,max} = 8.16 \times
10^{14}\,\mathrm{g\,cm}^{-3}$ is the critical density at which the
very last point in the sequence of decreasing $r_\mathrm{p/e}$ (the
one shown in Fig.~\ref{fig:mbDiffRot2}) jumps discontinuously to large
$r_\mathrm{e}$. At $\rho_\mathrm{b,max} = 9.21 \times
10^{14}\,\mathrm{g\,cm}^{-3}$, which is above the critical density for
c30p10 in Fig.~\ref{fig:mbDiffRot2}, the c30p10 models become
quasitoroidal as $r_\mathrm{p/e}$ decreases and $\Omega_c$
increases. They exhibit the same systematics as the c30p5 and cold
models. We note that what we have described for the c30p10 models also
occurs for the c30p5 models, although at significantly lower densities
$\rho_\mathrm{b,max} \lesssim 5\times10^{14}\,\mathrm{g\,cm}^{-3}$ and
even the cold models show similar trends at low densities.

The sequences shown in Figs.~\ref{fig:mbDiffRot1} and
\ref{fig:mbDiffRot2} are extreme configurations in the sense that
models with smaller $r_\mathrm{p/e}$ cannot be found by the CST solver
and may not exist for the rotation law that we consider here. Real
HMNS may not by such critical rotators. In
Fig.~\ref{fig:mVsOmegaDiffRot}, we plot $M_\mathrm{b}$ and
$M_\mathrm{g}$ for the LS220 EOS as a function of central angular
velocity $\Omega_c$ and temperature prescription. We fix the degree of
differential rotation to $\tilde{A}=1$ and show sequences in
$\Omega_c$ for a fixed maximum density $\rho_\mathrm{b,max} = 9.21
\times 10^{14}\,\mathrm{g}\,\mathrm{cm}^{-3}$, which is the highest
density shown in Fig.~\ref{fig:reVsOmegaDiffRot}. The transition to
quasitoroidal shape is smooth and quasitoroidal configurations are
marked with symbols. The end points of the $M_\mathrm{b}$ curves shown
in Fig.~\ref{fig:mVsOmegaDiffRot} and in
Fig.~\ref{fig:reVsOmegaDiffRot} correspond to the $M_\mathrm{b}$
values of the $\tilde{A}=1$ curves in Figs.~\ref{fig:mbDiffRot1} and
\ref{fig:mbDiffRot2} at $9.21 \times
10^{14}\,\mathrm{g}\,\mathrm{cm}^{-3}$.

Fig.~\ref{fig:mVsOmegaDiffRot} shows that, as in the case of uniform
rotation (cf.~Fig.~\ref{fig:omegaUniformRot}), hotter
\emph{subcritically} differentially spinning configurations have
higher $M_\mathrm{g}$. At the density chosen for this plot, they also
have higher $M_\mathrm{b}$, but at the higher densities at which the
masses of uniformly spinning models peak, the $M_\mathrm{b}$ of hotter
configurations are smaller than those of colder ones. It is
particularly remarkable that the models with the hot plateau at low
densities show the greatest thermal enhancement. They also transition
to a quasitoroidal shape last but terminate the earliest in
$\Omega_c$.  Nevertheless, for the $\rho_\mathrm{b,max}$ chosen here,
they can support slightly more mass at critical rotation than their
counterparts without low-density temperature plateau.

\vspace*{-0.3cm}
\section{Discussion and comparison with 3D NSNS simulations}
\label{sec:compare}
\vspace*{-0.2cm}

\subsection{The Stability of HMNS Equilibrium Sequences}
\label{sec:stability}
The existence of a maximum mass for equilibrium sequences of
nonrotating (TOV) neutron stars is one of the most important
astrophysical consequences of general relativity and, hence, is well
known in the study of compact objects.  The parameter space of hot
differentially rotating HMNS models studied here is vast and complex.
In the following, we briefly review the classical results on the
stability of stationary neutron stars and formulate how one may reason
regarding the stability of HMNS equilibrium models.

A particular useful approach to the stability problem is the
\textit{turning-point method} of \cite{sorkin:82}. The turning-point
method allows one to reason about the stability of sequences of
equilibrium solutions solely by examining the parameter space of
equilibrium models without dynamical simulations or linear
perturbation analysis.  The turning-point method has been used
extensively in previous work on the stability of cold
and uniformly rotating neutron stars (e.g., CST,
\citealt{friedman:88,stergioulas:95,read:09a}).

An equilibrium sequence is a one dimensional slice from the space of
equilibrium models indexed by some parameter. Here we use
$\rho_\mathrm{b,max}$ as our sequence parameter.  A model in the space
of equilibrium models may be defined by the following conserved
quantities: the gravitational mass $M_\mathrm{g}$, baryonic mass
$M_\mathrm{b}$, total angular momentum $J$, and total entropy $S$.
Generally, as one changes the sequence parameter,
$\rho_\mathrm{b,max}$, the quantities ($M_\mathrm{g}$, $M_\mathrm{b}$,
$J$, $S$) will vary.  A \textit{turning point} in the sequence occurs
when 3 out of 4 of the derivatives $d/d\rho_\mathrm{b,max}$ of
($M_\mathrm{g}$, $M_\mathrm{b}$,$J$, $S$) vanish.  For this point in
$\rho_\mathrm{b,max}$, the turning point theorem shows (i) that the
derivative of the fourth quantity in the tuple also vanishes, and (ii)
that the sequence must have transitioned from stable to unstable
(\citealt{sorkin:82} and \citealt{kaplan:14phd}).   This
characterization of the space of equilibrium models relies on the
assumption that the change in $M_\mathrm{g}$ depends to first order
only on the \textit{total} changes in baryonic mass $M_\mathrm{b}$,
angular momentum $J$, and entropy $S$, and not on changes to their
higher moments.  That is, changes in the distribution of entropy,
baryonic mass and angular momentum.  In nature, this will generally
not be the case, since cooling and angular momentum redistribution
will change the entropy and angular momentum distributions,
respectively. However, these changes will be slow and not drastic so
that changes to the total energy due to changes in these higher order
moments will be small. We account for such changes approximately by
considering different degrees of differential rotation and a range of
temperature prescriptions in the following.

If we are considering the special case of zero-temperature
configurations, then the entropy $S$ is no longer relevant to the
equilibrium's stability, since the change to the configuration's
energy due to a change in entropy is also zero. In this case, a
turning point may be identified when two out of three of the set
$d/d\rho_\mathrm{b,max}(M_\mathrm{g}, M_\mathrm{b}, J)$ are zero.
Zero temperature is a very good approximation for our cold equilibrium
models.  In Fig.~\ref{fig:mgVsRhoConstMb}, we plot $M_\mathrm{g}$
along constant $M_\mathrm{b}$ sequences with $M_\mathrm{b}=2.9
M_\odot$ for the HShen EOS ($M_\mathrm{b}=2.9 M_\odot$ corresponds
to $M_\mathrm{b}$ of a HMNS formed from two NSs of $M_\mathrm{g}=1.35
M_\odot$, assuming no mass loss).  All of these curves have a minimum
located at $\rho_\mathrm{b,max} \gtrsim 1 \times 10^{15}
\mathrm{g\,cm}^{-3}$.  

For the cold sequences, these minima are turning points because
$dM_\mathrm{g}/d\rho_\mathrm{b,max}$ and
$dM_\mathrm{b}/d\rho_\mathrm{b,max}$ are both zero. Any models along
those curves at densities in excess of $\rho_\mathrm{b,max}$ at the
minima are secularly unstable to collapse. For the hot temperature
parametrizations\footnote{We show only the c40p0 and cold temperature
  parametrizations in Fig.~\ref{fig:mgVsRhoConstMb}, because we find
  them to be the limiting cases. All other parametrizations have
  minima at intermediate locations in the ($M_\mathrm{g}$,
  $\rho_\mathrm{b,max}$) plane.}, the minima are only approximations
to the turning point (which we shall call \textit{approximate turning
  points}) because only two out of four
($dM_\mathrm{g}/d\rho_\mathrm{b,max}$ and
$dM_\mathrm{b}/d\rho_\mathrm{b,max}$) of the derivatives of
($M_\mathrm{g}$, $M_\mathrm{b}$, $J$, $S$) are zero.  We argue that
these approximate turning points are good indicators of the onset of
instability for the equilibrium sequences for several reasons.  (i) We
find that the approximate turning points for all considered
temperature parametrizations and measures of differential rotation
($\tilde{A} = 0$ to $\tilde{A} = 1.1$ with spacing $\delta\tilde{A} =
0.1$) lie within the same $\sim 25\%$ range in $\rho_\mathrm{b,max}$
indicated by the blue lines in Fig.~\ref{fig:mgVsRhoConstMb}
(similarly within a $\sim 25\%$ range in $\rho_\mathrm{b,max}$ for the
LS220 EOS).  (ii) In cold uniformly rotating NS models, approximate
turning points occur where one out of three of
$d/d\rho_\mathrm{b,max}(M_\mathrm{g}, M_\mathrm{b}, J)$ vanish. The
study of such models shows that the actual turning point density is
within only $\sim 1\%$ of the approximate turning point density (where
$dM_\mathrm{g}/d\rho_\mathrm{b,max} = 0$ along the mass-shed sequence;
cf. Fig.~10 of \citealt{stergioulas:95}).  (iii) The turning-point
condition is a sufficient, but not necessary, criterion for secular
instability. Thus instability must set in at $\rho_\mathrm{b,max}$
greater than the turning-point $\rho_\mathrm{b,max}$, but may set in
already at lower densities (see, e.g., \citealt{takami:11} for an
example).  It is thus conservative to use the approximate turning
point located at the highest $\rho_\mathrm{b,max}$ over all sequences
for a given EOS as an upper bound for the maximum stable
$\rho_\mathrm{b,max}$ of HMNS models for that EOS.

Further to the above, we have verified (see Sec.~9.1 of
\citealt{kaplan:14phd}) that the same density ranges contain
approximate turning points when examining alternate pairs of conserved
variables: both $J$ and $M_\mathrm{b}$, and $J$ and $M_\mathrm{g}$ (in
contrast to Fig.~\ref{fig:mgVsRhoConstMb}, where we examine
$M_\mathrm{b}$ and $M_\mathrm{g}$).  This gives us confidence that the
method of approximate turning points is self consistent with respect
to choice of the vanishing derivatives.  Unfortunately, since the CST
code employs only barytropic EOSs, we lack the infrastructure
necessary to study the total entropy of the configurations, and note
that an examination of total entropy of these models is an important
goal for future work.

\vspace*{-0.25cm}
\subsection{The Secular Evolution of HMNS from Mergers}
\label{sec:secularEvolution}
\vspace*{-0.1cm}

A HMNS remnant resulting from the merger of two NSs that does not
promptly collapse into a black hole will settle into a
quasiequilibrium state.  More precisely, this is a state in which the
HMNS is no longer in dynamical evolution, measured, for example, by
oscillations in the HMNS maximum density.  This should occur several
dynamical times after merger.  From this point on, the HMNS will
evolve secularly along some sequence of equilibrium models.  A secular
evolution is, by definition, a dissipative process that may involve
energy loss\footnote{ The trapped lepton number is, of
  course, also changing, since the fluxes of $\nu_e$ and $\bar{\nu}_e$
  will at least initially not be symmetric. However the effect of the
  trapped lepton fraction on stability is minimal, since electron
  degeneracy pressure is present only at high densities where it is
  much smaller than the baryon pressure in hot HMNSs
  that lose energy to neutrino emission (see
  Fig.~\ref{fig:pressure_contributions}).} from the
system. Consequently, we may parametrize the secular evolution of the
HMNS towards a turning point via the change in its total mass-energy,
which, in our case, is the change in gravitational mass of the
equilibrium model. This occurs in HMNSs via neutrino cooling and the
emission of gravitational radiation. In addition, the rotational
energy of the HMNS may be reduced by angular momentum redistribution
via the MRI, provided this occurs sufficiently slowly to be
characterized as as secular process. This can lead to a build up of
magnetic field, or dissipation of the free energy of differential
rotation as heat (see, e.g., \citealt{thompson:05} for a detailed
discussion), which may lead to increased neutrino
cooling. Furthermore, specific angular momentum transported to the
HMNS surface may unbind surface material, leading to a decrease in $J$
and $ M_\mathrm{b}$. These changes of $M_\mathrm{b}$ and
$J$ may be significant, but cannot be taken into account by the approximate
description of the HMNS's evolution we are considering here. Our results
should thus be interpreted with these limitations in mind.

\begin{figure}[t]
\centering
\includegraphics[width=\columnwidth]{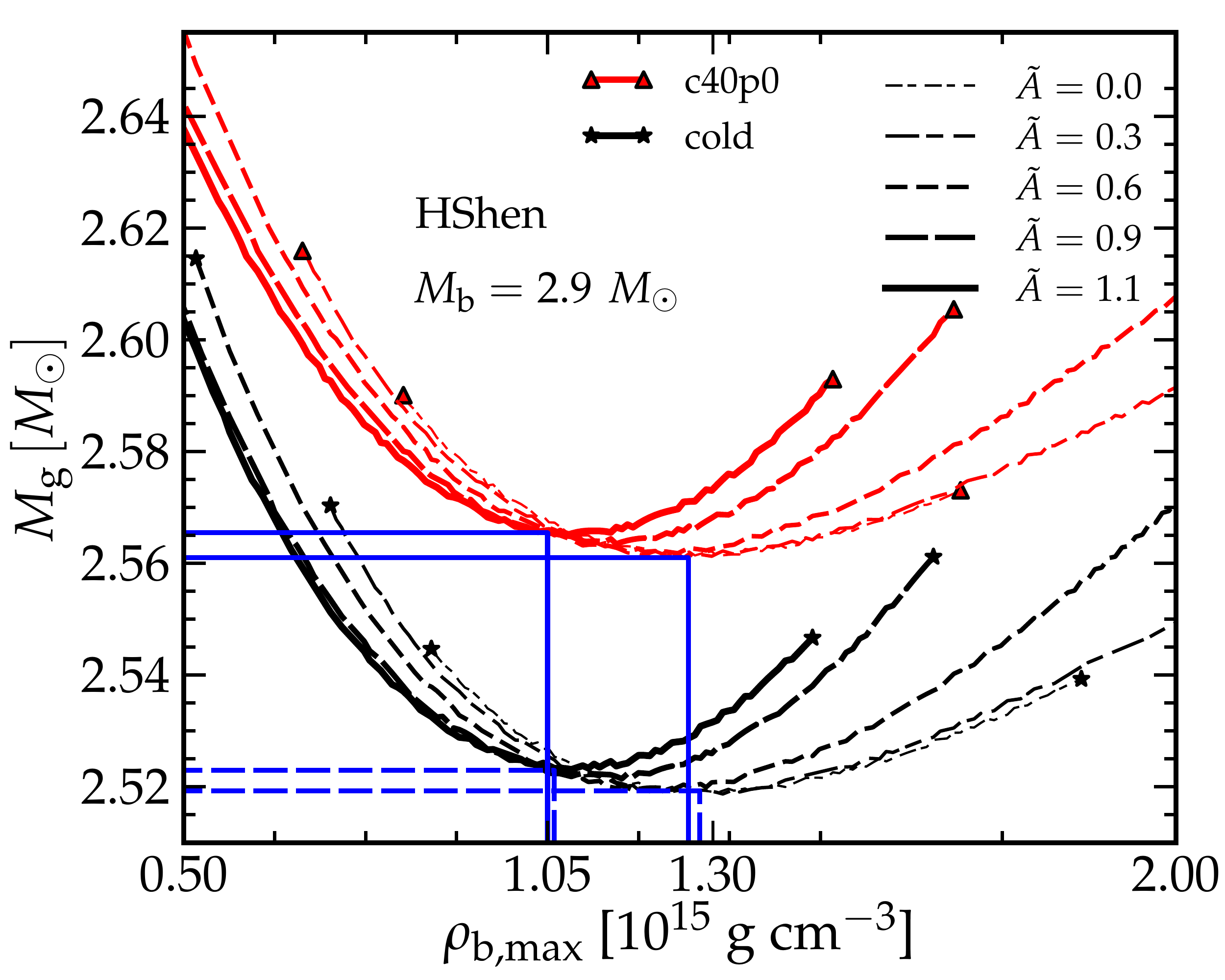}
\caption{Gravitational mass $M_\mathrm{g}$ as a function of maximum
  baryon density $\rho_{b,\mathrm{max}}$ for models with
  $M_\mathrm{b}=2.9 M_\odot$.  Each curve is for a fixed degree of
  differential rotation $\tilde{A}$, with the axis ratios $r_{p/e}$
  chosen such that $M_\mathrm{b}=2.9 M_\odot$.  Symbols mark
  equilibrium solutions at the minimum $r_{p/e}$ for which a solution
  can be found for $M_\mathrm{b}=2.9\,M_\odot$ and a given $\tilde{A}$
  (i.e., the solver fails to converge when searching for
    a $M_\mathrm{b}=2.9 M_\odot$ mass model at densities outside the
    bounds of the symbols).  The local minima of these curves are
  approximate turning points of the sequences.  For the cold (c40p0)
  models, we have noted the range in $M_\mathrm{g}$ and
  $\rho_{b,\mathrm{max}}$ across models with different amounts of
  differential rotation with dashed (solid) blue lines. Consequently,
  $\rho_{b,\mathrm{max}} = 1.30 \times 10^{15} \mathrm{g\, cm}^{-3}$
  represents the upper limit for the baryon density of a stable HMNS
  with the HShen EOS.  Note also that the difference in $M_\mathrm{g}$
  of the approximate turning points between sequences with the same
  temperature prescription is only $\sim 0.005 M_\odot$}
\label{fig:mgVsRhoConstMb}
\end{figure}

A secularly evolving HMNS will, in general, evolve in the direction of
decreasing gravitational mass $M_\mathrm{g}$ while (at least
approximately) conserving its total baryonic mass $M_\mathrm{b}$.
This results in an increasing density and compactness of the
star. Figure~\ref{fig:mgVsRhoConstMb} shows, for a fixed temperature
prescription and differential rotation parameter, that the
gravitational mass $M_\mathrm{g}$ of a sequence with fixed baryonic
mass $M_\mathrm{b} = 2.9\,M_\odot$ (using the HShen EOS; we find
qualitatively the same for the LS220) is decreasing with increasing
density. This continues until, $M_\mathrm{g}$ reaches a minimum at an
approximate turning point for $\rho_{\mathrm{b,max}} \gtrsim 1\times
10^{15} \,\mathrm{g\, cm}^{-3}$.  Here, $\delta M_\mathrm{g} = 0$, and
$\delta M_\mathrm{b}$ vanishes by our choice of a constant
$M_\mathrm{b}$ sequence.

The curves in Fig.~\ref{fig:mgVsRhoConstMb} are shown for constant
differential rotation parameter $\tilde{A}$. However, a HMNS of
$M_\mathrm{b} = 2.9 M_\odot$ is not necessarily constrained to a
specific curve.  One would expect the HMNS to evolve to neighboring
curves of less extreme differential rotation (decreasing $\tilde{A}$),
in accordance with its loss of angular momentum due to gravitational
waves and its redistribution of angular momentum due to other secular
processes. Nevertheless, consider the limit in which the HMNS is
constrained to a curve of constant $\tilde{A}$. Then it would evolve
secularly until reaching the curve's minimum.  At this point, any
further energy loss implies that the HMNS must either (a) secularly
evolve to a nearby equilibrium sequence with lower temperature or
lower degree of differential rotation and higher density (another
curve on the plot) \emph{or} (b) undergo collapse to a black hole.
Note that the densities at which the minimum occurs for different
$\tilde{A}$ and temperatures are remarkably close to each other. For
the sequences using the HShen EOS shown in
Fig.~\ref{fig:mgVsRhoConstMb}, the approximate turning points lie in
the range $1.05\times 10^{15} \,\mathrm{g\, cm}^{-3}
<\rho_{\mathrm{b,max}} < 1.30\times 10^{15} \,\mathrm{g\, cm}^{-3}$
for all considered $\tilde{A}$ and both shown temperature
prescriptions.  The constant-$M_\mathrm{b}$ curves for other
temperature parametrizations (c20p0, c30p0, c30p5, c30p10) are all
located in-between the curves for the c40p0 and cold cases shown.
Thus, we expect that the point of collapse for a HMNS will be marked
by its evolution to this density regime regardless of the
  temperature distribution of the model. 

From the above findings, we conclude that thermal effects have little
influence on the stability of HMNSs in rotational equilibrium against
gravitational collapse. However, our results do imply that thermal
support \textit{will} affect at what density the HMNS first settles to
its quasiequilibrium state.  The discussion in
\S\ref{sec:results_2Ddiff} and, in particular,
Fig.~\ref{fig:mVsOmegaDiffRot}, illustrates that at subcritical
rotation rates and densities significantly below those of the
approximate turning points, models with hot temperature profiles have
a larger $M_\mathrm{b}$ compared to models with cooler temperatures at
the same $\rho_{\mathrm{b,max}}$.  Thus a HMNS with greater thermal
support will reach a quasiequilibrium at a lower
$\rho_{\mathrm{b,max}}$, and thus have more energy to lose before it
can evolve to the critical density regime for collapse.  

While thermal effects may be important in setting the initial
conditions for the secular evolution of a HMNS, they appear to be of
little consequence to the stability of a HMNS in quasiequilibrium.
Once in a quasiequilibrium state, the energy lost by a HMNS during its
secular evolution is the most robust indicator for its progress
towards instability and collapse. Fig.~\ref{fig:mgVsRhoConstMb} shows
that this is true regardless of the degree of differential rotation of
the HMNS. For a fixed temperature parametrization, the difference in
$M_\mathrm{g}$ between different degrees of differential rotation is
at most $\sim$$0.005\,M_\odot$, corresponding to $\lesssim 10\%$ of the
total energy lost during the HMNS's secular evolution.

\subsection{Comparison with NSNS Merger Simulations}

\cite{sekiguchi:11nsns} conducted simulations of NSNS mergers using
the HShen EOS and included neutrino cooling via an approximate leakage
scheme.  They considered three equal-mass binaries with component
NS gravitational (baryonic) masses of $1.35 M_\odot$ ($1.45 M_\odot$),
$1.50 M_\odot$ ($1.64 M_\odot$), $1.60 M_\odot$ ($1.77 M_\odot$)
denoted as L, M, and H, respectively. The HMNS formed from their
high-mass binary collapses to a black hole within $\lesssim
9\,\mathrm{ms}$ of merger. The low-mass and the intermediate-mass
binaries, however, form hot ($T \sim 5-30\,\mathrm{MeV}$) spheroidal
quasiequilibrium HMNSs that remain stable for at least
$25\,\mathrm{ms}$, the duration of their postmerger simulations.

\cite{sekiguchi:11nsns} argue that thermal pressure support could
increase the maximum mass of HMNSs with $T\gtrsim20\,\mathrm{MeV}$ by
$20-30\%$. The results that we lay out in \S\ref{sec:TOVresults} and
\S\ref{sec:CSTresults} of our study suggest that it is not
straightforward to disentangle centrifugal and thermal effects for
differentially rotating HMNS. Our findings show that critically
spinning configurations (i.e., configurations at which the maximum
$M_b$ is obtained for a given $\tilde{A}$) of hot models do not lead
to an increase in the maximum supported baryonic mass by more than a
few percent and in most cases predict a lower maximum mass than in the
cold case. We find it more useful to consider the results of
\cite{sekiguchi:11nsns} in the context of the evolutionary scenario
outlined in \S\ref{sec:secularEvolution}.

\begin{figure}[t]
\centering
\includegraphics[width=\columnwidth]{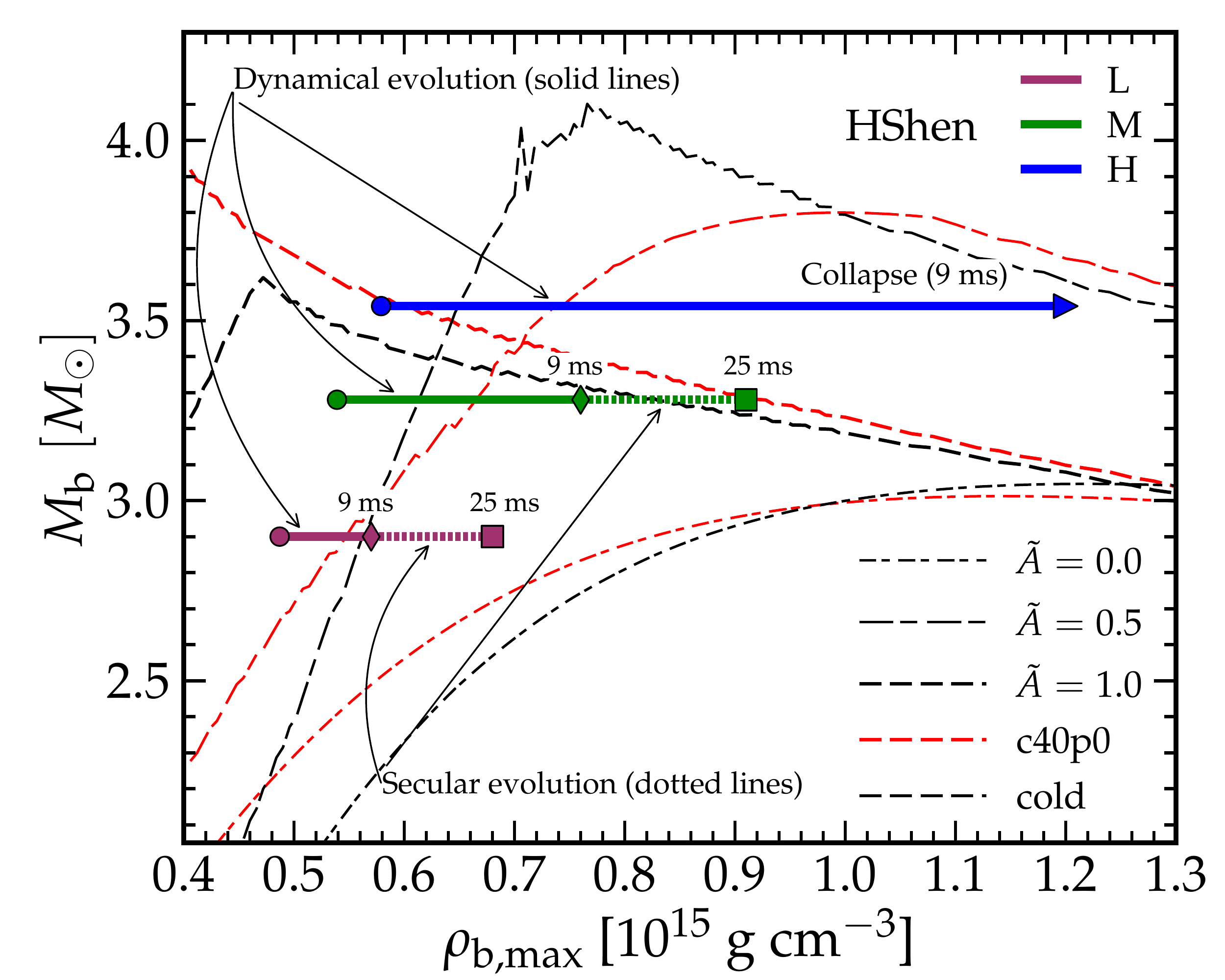}
\caption{Similar to Fig.~\ref{fig:mbDiffRot1} but for the HShen EOS
  and showing the approximate evolution of HMNSs from
\cite{sekiguchi:11nsns}.  We show the evolution of maximum density of
the HMNS for the low, medium and high mass configurations (thick lines
L, M and H) starting from the premerger density (noted by circles),
and ending at the simulation termination densities (squares, or, in
the H configuration, an arrow indicating collapse to a black hole).
After $\sim$$9\,\mathrm{ms}$ (noted with diamonds), the L and M models
show negligible dynamical oscillations and have settled to a
quasiequilibrium state.  From there until the end of the simulation,
the L and M HMNS are evolving secularly (indicated by thick dotted
lines).  We note that given the limitations of our
  approach discussed in the main text, the evolutionary tracks of
  constant baryonic mass shown in this figure should not be considered
  quantitatively reliable.}
\label{fig:diffRotNSNS}
\end{figure}

In Fig.~\ref{fig:diffRotNSNS}, we plot $M_\mathrm{b}$ as a function of
$\rho_\mathrm{b,max}$ for select sequences of uniformly and
differentially rotating models obtained with the HShen EOS with the
cold and c40p0 temperature prescriptions.  We also mark the immediate
postmerger densities of the L, M, and H models of
\cite{sekiguchi:11nsns} and their evolutionary tracks (in
$\rho_\mathrm{b,max}$). The high-mass model H never settles into a
quasiequilibrium and collapses to a black hole during the dynamical
early postmerger phase. Its $\rho_\mathrm{b,max}$ evolves within
$\sim$9\,ms from $0.58\times 10^{15}\,\mathrm{g\,cm}^{-3}$ to values
beyond the range of the plot. Our secular-evolution approach cannot be
applied to this model since it never reaches a quasiequilibrium state.
The lower-mass M and L models enter Fig.~\ref{fig:diffRotNSNS} at
successively lower densities. Their ``ring-down'' oscillations are
damped by $\sim$9\,ms after which the HMNSs evolve secularly with
$\rho_\mathrm{b,max}$ that increase roughly at the same rate in both
models, suggesting that their rate of energy loss is comparable.  At
such early times, gravitational waves are most likely dominating
energy loss (cf.\ the discussion of timescales in
\citealt{paschalidis:12}), and, indeed, model M and L exhibit similar
gravitational wave amplitudes and frequencies
(\citealt{sekiguchi:11nsns}, Fig.~4). Focusing on model L, we now
consider Fig.~\ref{fig:mgVsRhoConstMb}, which shows sequences of
constant $M_\mathrm{b}$ (for model L with $M_\mathrm{b} \sim
2.9\,M_\odot$). As the HMNS loses energy, $M_\mathrm{g}$ decreases and
the HMNS evolves to the right (towards higher
$\rho_\mathrm{b,max}$). Model L enters its secular evolution at a
central density of $\sim$$0.56\times 10^{15}\,\mathrm{g\,cm}^{-3}$ and
evolves secularly to $\sim$$0.68\times 10^{15}\,\mathrm{g\,cm}^{-3}$
within $\sim$$16\,\mathrm{ms}$. Largely independent of its specific
angular momentum distribution and thermal structure,
Fig.~\ref{fig:mgVsRhoConstMb} suggests that this model will reach its
global minimum $M_\mathrm{g}$ and, thus, instability in a small
density range of $\sim 1.05-1.30\,\times
10^{15}\,\mathrm{g\,cm}^{-3}$. 

Using our approximate secular evolution model for HMNSs discussed in
\S\ref{sec:secularEvolution}, we linearly extrapolate the density
evolution of model L in \cite{sekiguchi:11nsns}. We expect a possible
onset of collapse at $t \gtrsim 58\,\mathrm{ms}$ after merger (and
$\gtrsim$$49\,\mathrm{ms}$ after the start of the secular evolution).
These numbers should be regarded as very rough estimates, given the
limitations and rather qualitative nature of our model.  Depending on
its angular momentum when entering its secular evolution, its cooling
rate, angular momentum redistribution and loss, model L may
alternatively evolve into a long-term stable supramassive neutron
star, since a baryonic mass of $\sim$$2.9\,M_\odot$ can in principle
be supported by the HShen EOS at the supramassive limit
(cf.~Table~\ref{tab:uniform}).  Furthermore, we have also checked
  that model L of \cite{sekiguchi:11nsns} contains sufficient angular
  momentum to be represented by the sequences identified in
  Figs.~\ref{fig:mgVsRhoConstMb} and \ref{fig:diffRotNSNS}.  At a time
  of $\sim$$10-15$ ms after merger, model L has an angular momentum of
  $6\times10^{49}$ g cm$^2$ s$^{-1}$ (~6.8 in $c = G = M_{\sun}$
  units). Plots of
  similar sequences can be found in \cite{kaplan:14phd}.  They are
  consistent with this value.

The role of thermal pressure effects in all of the above
is relatively minor (cf.~the very similar $\rho_\mathrm{b,max}$
locations of the $M_\mathrm{g}$ minima in hot and cold configurations
shown in Fig.~\ref{fig:mgVsRhoConstMb}). However, when first entering
the secular regime as a subcritical HMNS, a configuration with higher
temperature and stronger thermal pressure support will be less compact
and will have a lower $\rho_\mathrm{b,max}$ at a fixed $M_\mathrm{b}$
than a colder one. Hence, in the picture of secular HMNS evolution
discussed in \S\ref{sec:secularEvolution}, such a configuration would
have to evolve ``farther'' in $\rho_\mathrm{b,max}$ to reach
criticality and, thus, can survive longer at fixed energy loss rates.

\cite{paschalidis:12} performed NSNS merger simulations of $\Gamma =
2$ polytropes in which they approximated a thermal pressure component
with a $\Gamma=2$ $\Gamma$-law. Their postmerger HMNS enters its
secular evolution in a quasitoroidal configuration with two
high-density, low-entropy cores, a central, lower-density, hot region
and a high-entropy low-density envelope. The total mass of their model
can be arbitrarily rescaled, but in order to estimate temperatures and
thermal pressure contributions, the authors scaled their HMNS remnant
to a gravitational mass of $2.69\,M_\odot$. With this, they estimated
in their quasitoroidal HMNS peak and rms temperatures of
$\sim$$20\,\mathrm{MeV}$ and $\sim$$5\,\mathrm{MeV}$,
respectively. \cite{paschalidis:12} studied the effect of neutrino
cooling on the HMNS evolution by introducing an ad-hoc cooling
function that removes energy proportional to the thermal internal
energy (neglecting the stiff temperature dependence of neutrino
cooling). In order to capture effects of cooling during the limited
simulated physical postmerger time, they drained energy from their
HMNS at rates $\sim$$100-200$ times higher than realistic cooling by
neutrinos.

The authors considered cases without cooling and with two different
accelerated cooling timescales.  As cooling is turned on in their
simulations, the slope of the maximum baryon density
$\rho_\mathrm{b,max}(t)$ of the HMNS increases discontinuously and the
higher the cooling rate, the faster the evolution to higher
$\rho_\mathrm{b,max}(t)$. The HMNSs in both cases with cooling become
unstable at different times, but roughly at the same
$\rho_\mathrm{b,max}$. This is consistent with the secular HMNS
evolution picture laid out in \S\ref{sec:secularEvolution}. Cooling
reduces the total energy of the system ($M_\mathrm{g}$) and drives the
HMNS to higher $\rho_\mathrm{b,max}$ at fixed $M_\mathrm{b}$ until the
(approximate) turning point is reached and collapse ensues. However,
losses due to gravitational wave emission and angular momentum
redistribution and shedding will have the same effect and may dominate
in nature, since they are likely to operate more rapidly than neutrino
cooling (cf.\ the discussion of timescales by
\citealt{paschalidis:12}).

\cite{bauswein:10} carried out smoothed-particle hydrodynamics
simulations of HMNSs in the conformal-flatness approximation to
general relativity.  They compared simulations using the full
temperature dependence of the HShen and LS180 EOS\footnote{The LS180
  is the variant of the \cite{lseos:91} EOS with nuclear
  compressibility modulus $K_0 = 180\,\mathrm{MeV}$.} with an
approximate treatment of thermal pressure via a $\Gamma$-law,
$P_\mathrm{th}=(\Gamma_\mathrm{th}
-1)\epsilon_\mathrm{th}\rho_\mathrm{b}$. Although \cite{bauswein:10}
do not provide a figure showing the evolution of maximum baryon
density, they show (in their Fig.~5) graphs of cumulative mass as a
function of distance from the center of the LS180-EOS HMNS at
$8\,\mathrm{ms}$ after merger, roughly the time when the dynamical
early postmerger phase is over and the secular HMNS evolution
begins. From this, it may be observed that the HMNS with the lower
thermal $\Gamma$ ($\Gamma_\mathrm{th} = 1.5$) is more compact than the
model with $\Gamma_\mathrm{th} = 2$. The HMNS evolved with the fully
temperature-dependent LS180 EOS is in between the two, but closer to
the $\Gamma_\mathrm{th} = 2$ model. \cite{bauswein:10} found that the
more compact HMNS with $\Gamma_\mathrm{th}=1.5$ collapses after 10 ms,
while the less compact $\Gamma_\mathrm{th}=2.0$ and full-LS180 cases
collapse after $\sim$$20\,\mathrm{ms}$. This is consistent with the
picture of secular HMNS evolution drawn in
\S\ref{sec:secularEvolution}: Given a fixed number of baryons, a less
compact configuration has a lower maximum baryon density after merger
and, therefore, begins its secular evolution (in the sense of
Figs.~\ref{fig:mgVsRhoConstMb} and \ref{fig:diffRotNSNS}) at a lower
density than a more compact configuration. Consequently, it must lose
more energy before reaching the critical density for collapse.

The above illustrates how thermal pressure effects may increase the
lifetime of a HMNS by affecting the initial conditions for its
secular evolution.  From \S\ref{sec:CSTresults} one notes that hot
configurations, at densities below $\lesssim
10^{15}\,\mathrm{g\,cm}^{-3}$ (the exact value being EOS dependent),
may support significantly larger masses than their cold counterparts
at the same $\rho_{b,\mathrm{max}}$.  Thus, during the dynamical
settle-down of two merging neutron stars to a secularly-evolving HMNS
remnant, a configuration with lower thermal pressure will need to
evolve to higher $\rho_{b,\mathrm{max}}$ to reach an equilibrium
configuration.

\section{Summary and Conclusions}
\label{sec:conclusions}

The merger of double neutron stars with component masses in the most
commonly observed mass range ($\sim$$1.3-1.4\,M_\odot$;
\citealt{lattimer:12}) is most likely to result in a hot,
differentially spinning hypermassive neutron star (HMNS) remnant that
is stable against collapse on a dynamical timescale, but likely
secularly evolving towards instability, driven by energy loss. While
a number of merger simulations in approximate or full
general-relativity with the necessary microphysics are now available,
the role of thermal pressure support on the postmerger HMNS and its
stability is not well understood.

In this study, we have attempted to gain insight into the role of
thermal pressure support by constructing nonrotating, uniformly
rotating and differentially rotating axisymmetric equilibrium
solutions with multiple microphysical, fully temperature and
composition dependent equations of state (EOS) and parametrized
temperature distributions motivated by results from full merger
simulations.  Such axisymmetric equilibrium models may be acceptable
approximations to merger remnants that have survived the initial
highly dynamical and strongly nonaxisymmetric postmerger evolution and
have settled down into longer-term stable quasiequilibrium. How far
away the equilibrium configurations really are from real HMNSs, and
the reliability of our results, will ultimately have to be established
by more detailed comparisons with merger simulations in future work.

In the secular postmerger phase, the baryonic mass $M_\mathrm{b}$ of
the hypermassive merger remnant is approximately conserved. Thus the
dependence of the maximum of $M_\mathrm{b}$ on temperature is the most
interesting quantity to study. In spherical symmetry (the TOV case),
we find that at densities significantly lower than the density at
which the maximum mass configuration occurs, thermal enhancement of
the NS mass can be strong. Generally, hotter configurations yield the
same $M_\mathrm{b}$ at lower central densities than their colder
counterparts. However, when considering compact maximum-$M_\mathrm{b}$
configurations, thermal effects are small.  For reasonable temperature
prescriptions, hot temperatures lead to a small ($\lesssim 1\%$)
decrease of $M_\mathrm{b}^\mathrm{max}$ for five out of the seven EOS
that we consider. The two other EOS, the HShen EOS and the
GShen-FSU2.1 EOS, show up to $\sim$$2\%$ thermal enhancement of
$M_\mathrm{b}$. As expected, none of the considered EOS could support
a remnant of the merger of a canonical double NS system with typical
masses.
 
Rapidly uniformly spinning configurations can support supramassive
NSs. We have studied uniformly spinning sequences generated with the
LS220 and HShen EOS. As in the TOV case, we find significant thermal
enhancement of $M_\mathrm{b}$ at low central densities and rotation
rates up to mass shedding. At high densities, however, thermal
pressure is much less important for the support of the inner NS core,
but bloats the envelope.  This results in hotter configurations
reaching mass shedding at lower angular velocities than colder
configurations. Hence, at the mass-shedding supramassive limit,
$M_\mathrm{b}$ and $M_\mathrm{g}$ decrease with increasing temperature
for uniformly spinning NSs. For the LS220 EOS (HShen EOS), the cold
supramassive $M_\mathrm{b}$ limit is $\sim$$2.823\,M_\odot$
($\sim$$3.046\,M_\odot$). Under the plausible assumption that the HMNS
merger remnant evolves towards a uniformly rotating configuration,
assuming no mass loss during or after merger, the cold supramassive
limit corresponds to component gravitational masses in an equal-mass
progenitor binary of $M_\mathrm{g} \sim 1.287\,M_\odot$ ($M_\mathrm{g}
\sim 1.403\,M_\odot$). On the other hand, a supramassive LS220 (HShen)
NS with a $30$-MeV core and a $10$-MeV envelope has a supramassive
limit $M_\mathrm{b} \sim 2.587\,M_\odot$ ($M_\mathrm{b} \sim
2.808\,M_\odot$), which corresponds to binary component $M_\mathrm{g}
\sim 1.185\,M_\odot$ ($M_\mathrm{b} \sim 1.300\,M_\odot$). Hence, cold
maximally uniformly rotating configurations of LS220 and HShen NSs may
barely support the merger remnant of canonical double NS binaries, but
hot ones might not.

Differential rotation adds yet another layer of complexity, but is the
most interesting scenario, since hypermassive merger remnants are born
with differential rotation. The notion of a maximum mass of a
differentially rotating HMNS is somewhat misleading, since different
rotation laws will give different masses and different solvers may
converge to different branches in the solution space. Hence, all
``maximum'' masses quoted are lower limits. For the commonly used
$j-const.$ rotation-law, parametrized by the dimensionless parameter
$\tilde{A}$, we find $M_\mathrm{b}$ up to $\sim 3.65\,M_\odot$ and
$\sim 4.10\,M_\odot$, for the LS220 EOS and the HShen EOS,
respectively. These high-mass configurations generally occur at
densities that are up to a factor of two lower than those of
maximum-$M_\mathrm{b}$ TOV and uniformly rotating models. Even higher
masses could be found, but such configurations would be dynamically
nonaxisymmetrically unstable.

Our results indicate that the role of thermal effects depends very
much on the degree of differential rotation in addition to maximum
density and (central) angular velocity. All qualitative findings are
identical for the LS220 EOS and the HShen EOS. For critically rotating
models (with minimum axis ratio $r_\mathrm{p/e}$ for which a solution
is found) the dependence on differential rotation is as follows: (i)
For a low degree of differential rotation ($\tilde{A} \lesssim 0.4$),
the same systematics as found for the uniformly rotating case
hold. (ii) In models with intermediate degree of differential rotation
($\tilde{A} \sim 0.5-0.7$), hot configurations have systematically
lower ``maximum'' $M_\mathrm{b}$ than colder ones. (iii) Models with
high degree of differential rotation ($\tilde{A} \gtrsim 0.7$) are
mostly quasitoroidal and the ``maximum'' $M_\mathrm{b}$ occurs at low
densities ($\lesssim 5\times 10^{14}\,\mathrm{g\,cm}^{-3}$) and is
mildly enhanced by thermal pressure support for models with hot cores,
but cold envelopes. Models with high-temperature envelopes remain
spheroidal until higher densities and have lower ``maximum''
$M_\mathrm{b}$.  The situation is yet different for differentially
rotating configurations that are rotating rapidly, but
subcritically. For example, for LS220 EOS configurations with
$\tilde{A} = 1$, models with thermally supported envelopes have the
highest $M_\mathrm{b}$ at subcritical rotation, but their sequences
terminate at lower angular velocities (higher $r_\mathrm{p/e}$) than
the cold configuration, which ultimately catches up in $M_\mathrm{b}$
at critical rotation.

To summarize all of the above: The forecast is mixed -- the role of
thermal effects on the baryonic mass that is supported by a given
configuration depends sensitively and in a complicated way on its
details, that is, central/mean baryon density, temperature
distribution, degree of differential rotation and rotation rate, to
name the most important parameters. Configurations that yield
``maximum'' $M_\mathrm{b}$ are essentially unaffected by thermal
effects.  Beyond that, no simple general statements can be made.

A more useful way to reason about the role of thermal pressure support
is to consider evolutionary sequences of equilibrium models
representing the secular quasiequilibrium evolution of a HMNS.  This
evolution occurs along tracks of approximately constant
baryonic mass $M_\mathrm{b}$ parametrized by maximum baryon density
$\rho_\mathrm{b,max}$. Since energy is lost by gravitational wave and
neutrino emission, a configuration always evolves into the direction
of decreasing total energy (i.e.\ decreasing gravitational mass
$M_\mathrm{g}$ and increasing $\rho_\mathrm{b,max}$). The turning
point theorem \citep{sorkin:82,friedman:13} says that an extremum in
$M_\mathrm{g}$ may mark the point at which the sequence becomes
secularly unstable to collapse.  While this can be proven rigorously
only for uniformly rotating (or nonrotating) configurations, we
conjecture that it also holds at least approximately for the much more
complex HMNS case. Provided this is true, we can define approximate
turning points using constant-$M_\mathrm{b}$ sequences with different
degrees of differential rotation and temperature
parametrizations. With this, we find that the approximate turning
points for a given $M_\mathrm{b}$ always lie in narrow ranges of
$\rho_\mathrm{b,max}$ and $M_\mathrm{g}$, which define the
$M_\mathrm{g}-\rho_\mathrm{b,max}$ space in which collapse to a black
hole occurs. Furthermore, the approximate turning point density at
which collapse must set in depends only very weakly on temperature.
Finally, we note that all approximate turning points found in this
work are at baryon densities below the cricial value for stable TOV
stars. This may suggest that HMNS with maximum densities at or higher
than the critical TOV central density could always be unstable to
collapse. This possibility should be investigated further in future
work.

Under the assumptions of the model laid out in this paper, the secular
evolution of a HMNS can then be described by the progressive decrease
of its gravitational mass $M_\mathrm{g}$ and increase of its maximum
density $\rho_\mathrm{b,max}$. Our results show that a HMNS with more
thermal pressure support will enter its secular evolution at a higher
$M_\mathrm{g}$ and lower $\rho_\mathrm{b,max}$ than a colder one (with
the same rotational setup). Hence, the hot HMNS will have to evolve
further in $\rho_\mathrm{b,max}$ until reaching its approximate
turning point. This explains the effects of thermal pressure observed
in merger simulations (e.g.,
\citealt{bauswein:10,sekiguchi:11nsns}). We note that the same
argument may also be applied to differences in HMNS spin: a more
rapidly spinning HMNS will enter its secular evolution at lower
$\rho_\mathrm{b,max}$ and higher total energy and, hence, will have to
evolve further in $\rho_\mathrm{b,max}$ to reach its approximate
turning point.

The goal of the work presented in this paper was to elucidate the role
of thermal pressure support in hypermassive NSNS merger remnants on
the basis of stationary spherically symmetric and axisymmetric
equilibrium solutions of the Einstein-Euler equations. While yielding
new insights, our present approach is limited in multiple ways:
(\emph{i}) Even in the secular quasiequilibrium evolution phase, HMNS
are not exactly axisymmetric. The CST solver used in this study does
not support nonaxisymmetric configurations, which makes it impossible
for us to test how sensitive our results are to symmetry assumptions.
(\emph{ii}) The equilibrium sequences considered here rely on an
ad-hoc rotation law and ad-hoc temperature and composition
parametrizations motivated by the simulations of
\cite{sekiguchi:11nsns}. In general, the angular velocity distribution
will be more complex (see, e.g., \citealt{galeazzi:12}) and the
temperature and composition of a HMNS will not be 
single-parameter functions of density. (\emph{iii}) The CST solver has
difficulties converging for configurations with a high degree of
differential rotation and it is not clear if the terminating axis
ratio $r_\mathrm{p/e}$ is set by the formulation and implementation of
the equations by the CST solver or if the termination occurs for
physical reasons. This could be checked only by a comparison study
with a more robust solver, e.g., the one of \cite{ansorg:03}.
(\emph{iv}) The approximate turning point theorem that we have used to
reason about the evolution and stability of HMNSs is heuristic and
lacks rigorous foundation. Fully reliable statements about the
stability of differentially rotating HMNSs with complex temperature
and compositional distributions will require at least perturbative
stability analysis or direct non-linear simulation.

Future work should address the above limitations (\emph{i}-\emph{iv})
and should also consider rotating configurations constructed with a
broader set of finite-temperature microphysical equations of state.

\section*{Acknowledgements}

We thank Eliot Quataert for inspiration and acknowledge helpful
discussions with Lars Bildsten, Ursula C.~T.~Gamma, Jim Lattimer, Lee
Lindblom, Sterl Phinney, Jocelyn Read, Yuichiro Sekiguchi, Masaru
Shibata, Saul Teukolsky, Kip Thorne, and especially Aaron
Zimmerman. Furthermore, we thank the anonymous referee for suggestions
that improved this paper. This work was initiated at a Palomar
Transient Factory Theory Network meeting at the Sky House, Los Osos,
CA. CDO wishes to thank the Yukawa Institute for Theoretical Physics
for hospitality during the long-term workshop Gravitational Waves and
Numerical Relativity 2013 when this work was completed. This research
is supported in part by NASA under the Astrophysics Theory Grant
no.~NNX11AC37G, by the National Science Foundation under grant
nos.~AST-1205732, PHY-1151197, AST-1212170, PHY-1068881, and
PHY-1068243, by the Alfred P. Sloan Foundation, and by the Sherman
Fairchild Foundation.  The calculations underlying the results
presented in this paper were performed on the Caltech compute cluster
``Zwicky'' (NSF MRI award No.\ PHY-0960291). The EOS tables, driver
and TOV solver routines used in this work are available for download
at {\tt http://www.stellarcollapse.org}. The solver for axisymmetric
equilibrium configurations of rotating HMNSs is not open source, but a
similar solver may be obtained form {\tt http://www.lorene.obspm.fr/}.
The figures in this paper were generated with the {\tt matplotlib}
library for {\tt Python} \citep{matplotlib}.

\appendix
\section{Temperature Parametrizations}
\label{app:temp}

\begin{deluxetable}{lcccc}
  \tablecolumns{5} \tablewidth{0pc} \tablecaption{Temperature Prescription Parameters} 
\tablehead{ 
Model & $T_\mathrm{max}$ &  Midpoint $m$ &  Scale $s$ & Plateau Temperature $T_p$
\\ 
& [MeV] & $\log_{10}(\rho_\mathrm{b} [\mathrm{g\,cm}^{-3}])$ & $\log_{10}(\rho_\mathrm{b} [\mathrm{g\,cm}^{-3}])$ & [MeV]
}
\startdata 
cold   & $-$ & $-$ &  $-$  & $-$  \\
c20p0  & 20 & $14.0\phantom{575} - 0.07$ & $0.25\phantom{25} $ & \phantom{1}0 \\
c30p0  & 30 & $14.125\phantom{5} - 0.07$ & $0.375\phantom{2}$ & \phantom{1}0 \\
c30p5  & 30 & $14.1875 - 0.07$ & $0.3125$ & \phantom{1}5 \\
c30p10 & 30 & $14.25\phantom{75} - 0.07$ & $0.25\phantom{25} $ & 10 \\
c40p0  & 40 & $14.25\phantom{57} - 0.07$ & $0.5\phantom{752} $& \phantom{1}0 
\enddata
\tablecomments{Parameters used for the temperature parametrizations
  used in this study.  The notation is c<core temperature>p<plateau
  temperature>.  All low-density temperature plateaus are tapered off
  at densities below $\sim$$10^{12}\,\mathrm{g\,cm}^{-3}$ with a tanh
  function with a midpoint at $\log_{10}(\rho_\mathrm{b}
  [\mathrm{g\,cm}^{-3}])=11.5$ and an $e$-folding width of
  $\log_{10}(\rho_\mathrm{b} [\mathrm{g}\,\mathrm{cm}^{-3}])=0.25$.  All
  minimum temperatures are $0.01\,\mathrm{MeV}$.  See
  Fig. \ref{fig:tempScripts} for a comparison of the various
  temperature prescriptions. The functional form of the prescriptions
  is given by (\ref{eq:hotCoreTofRho}) and
  (\ref{eq:plateauTofRho2}). }
\label{tab:tempScriptParams}
\end{deluxetable}

We consider temperature prescriptions with only a hot core at and
above nuclear density and with a hot core and a more extended
high-density plateau at lower densities. We emphasize that these
prescriptions are rather ad-hoc and motivated primarily by the data
from the simulations of \cite{sekiguchi:11nsns}. All high-temperature
regions are smoothly tapered-off (``rolled-off'') using tanh
functions.

The prescriptions with only a hot core (i.e. prescriptions cXp0) are
given by the 
\begin{equation}
T_\mathrm{roll}(\rho_\mathrm{b};T_{{1}},T_{{2}},m,s) = T_{{2}} + \frac{\left( T_{{1}} - T_{{2}} \right)}{2} \left(\tanh{\frac{\left(\log_{10}(\rho_\mathrm{b}) - m\right)}{s}} + 1 \right),
\label{eq:hotCoreTofRho}
\end{equation}
where $m$ is the roll-off midpoint (in $\log_{10}(\rho_\mathrm{b}
[\mathrm{g}\,\mathrm{cm}^{-3}]$) and $s$ is the roll-off $e$-folding
scale (also in $\log_{10}(\rho_\mathrm{b}
[\mathrm{g}\,\mathrm{cm}^{-3}]$).  For prescriptions that only have
hot cores, $T_{1}$ is set to the peak temperature $T_\mathrm{max}$ and
$T_{2}$ is set to $T_{\mathrm{min}} = 0.01\,\mathrm{MeV}$.  The
prescriptions with a high-temperature plateau at lower densities,
i.e.\ c30p5 and c30p10, are constructed as the sum of two of the above
functions as follows:

\begin{equation}
T(\rho_\mathrm{b};T_{\mathrm{max}},T_\mathrm{min}, T_{p},m',s') = T_\mathrm{min} \,+\, T_\mathrm{roll}(\rho_\mathrm{b};T_1=T_p,T_2 = 0, m=11.5,s=0.25) \,+\, T_\mathrm{roll}(\rho_\mathrm{b}; T_{1} = T_\mathrm{max}-T_{p}, T_2 = 0, m', s'),
\label{eq:plateauTofRho1}
\end{equation}
where $m'$ is the roll-off midpoint, $s'$ is the roll-off scale, and
$T_p$ is the plateau temperature. Writing this out more explicitly, we
have:

\begin{equation}
T(\rho_\mathrm{b};T_\mathrm{max},T_\mathrm{min},T_p,m,s) = T_\mathrm{min} + \frac{ T_p }{2} \left(\tanh{\frac{\left(\log_{10}(\rho_\mathrm{b}) - 11.5\right)}{0.25}} + 1 \right) + \frac{ T_\mathrm{max} - T_p }{2} \left(\tanh{\frac{\left(\log10(\rho_\mathrm{b}) - m\right)}{s}} + 1 \right)
\label{eq:plateauTofRho2}.
\end{equation}

Table~\ref{tab:tempScriptParams} summarizes the parameters 
for generating the temperature prescriptions used in this study.

\section{Solving for the Electron Fraction}
\label{sec:ye}
For a given EOS and temperature prescription, we find the electron
fraction $Y_e$ by first solving for $Y_e$ assuming neutrino-less
$\beta$-equilibrium for the cold case ($T=0.01\,\mathrm{MeV}$ or the
lowest temperature point available in the EOS table), using the
condition
\begin{equation}
\label{eq:nlbeta}
\mu_\nu = 0 = \mu_n + \mu_p - \mu_e\,\,,
\end{equation}
for the chemical potentials. In the absence of neutrinos, the lepton
fraction $Y_\mathrm{lep} = Y_e$.  In the hot case, neutrinos are
trapped in the HMNS matter above $\rho = \rho_\mathrm{trap} \approx
10^{12.5}\,\mathrm{g\,cm}^{-3}$ and $Y_\mathrm{lep} = Y_e + Y_\nu$,
where $Y_\nu = Y_{\nu_e} - Y_{\bar{\nu}_e}$.

We then take $Y_\mathrm{lep}$ and solve for $Y_e$ in the hot case with
neutrinos by treating the latter as a relativistic Fermi gas in
equilibrium for which $Y_\nu$ can be calculated from the neutrino
number density $n_\nu = n_{\nu_e} - n_{\bar{\nu}_e}$ via
\begin{equation}
Y_\nu = \frac{n_\nu}{\rho N_A}\,\,.
\label{eq:ynu1}
\end{equation}

The neutrino number density is 
\begin{equation}
n_\nu = 4\pi \left(\frac{k_B T}{hc}\right)^3 \left[ F_2(\eta_\nu) - F_2(-\eta_\nu)\right]\,,
\label{eq:ynu2}
\end{equation}
where $\eta_\nu = \mu_\nu / (k_B T)$ is the neutrino degeneracy
parameter \citep{bludman:78}. Note that in equilibrium, $\nu_e$ and
$\bar{\nu}_e$ have equal and opposite chemical potentials. $F_2$
is a Fermi integral given by
\begin{equation}
F_k(\eta) = \int_0^\infty \frac{x^k dx}{e^{x-\eta} +1}\,\,.
\end{equation}
In practice, we use 
\begin{equation}
F_2(\eta) - F_2(-\eta) = \frac{1}{3}\eta(\eta^2 + \pi^2)\,\,,
\label{eq:ynu3}
\end{equation}
which is given in \cite{bludman:78} and is exact for any degeneracy
parameter $\eta$.

We find $Y_e$ by finding the root
\begin{equation}
0 = Y_\mathrm{lep} - (Y_e + Y_\nu)\,\,.
\end{equation}
$Y_\mathrm{lep}$ is a fixed input. We set $Y_e = Y_\mathrm{lep}$ as an
initial guess and $Y_\nu$ is calculated using Eqs.~(\ref{eq:ynu1}),
(\ref{eq:ynu2}), and (\ref{eq:ynu3}), with $\mu_\nu = \mu_n + \mu_p -
\mu_e$ obtained from the EOS. $Y_e$ is then adjusted and we iterate
until convergence.

Since neutrinos begin to stream freely below $\rho_\mathrm{trap}$, we
also compute $Y_e$ using the $\nu$-less $\beta$-equilibrium condition
(Eq.~\ref{eq:nlbeta}). We then compute a final effective $Y_e$ using
\begin{equation}
Y_{e,\mathrm{eff}}(\rho,T[\rho]) = Y_{e,\nu-\mathrm{less} \,\beta}  (\rho,T[\rho])\, \times\,  (1-e^{-\rho_\mathrm{trap}/\rho}) + 
Y_{e,\beta}(\rho,T[\rho])\, \times\, e^{-\rho_\mathrm{trap}/\rho}\,\,.
\end{equation}

\end{document}